# A Pliable Index Coding Approach
# to Data Shuffling

Linqi Song, Christina Fragouli, and Tianchu Zhao

### Abstract

A promising research area that has recently emerged, is on how to use index coding to improve the communication efficiency in distributed computing systems, especially for data shuffling in iterative computations. In this paper, we posit that pliable index coding can offer a more efficient framework for data shuffling, as it can better leverage the many possible shuffling choices to reduce the number of transmissions. We theoretically analyze pliable index coding under data shuffling constraints, and design a hierarchical data-shuffling scheme that uses pliable coding as a component. We find benefits up to $O(ns/m)$ over index coding, where $ns/m$ is the average number of workers caching a message, and $m$, $n$, and $s$ are the numbers of messages, workers, and cache size, respectively.

## I. Introduction

A promising research area that has recently emerged, is on how to use coding techniques to improve the communication efficiency in distributed computing systems [2], [3], [4]. In particular, index coding has been proposed to increase the efficiency of data shuffling, that can form a major communication bottleneck for big data applications [2], [4], [5]. In index coding, a server has $m$ messages, and is connected through a broadcast channel to $n$ nodes; each node has a specific request, as well as some side information. The goal is to minimize the number of broadcast transmissions so that, each of the $n$ nodes receives its request. In this paper, we posit that using a form of pliable index coding, a variation of the traditional index coding, can offer a more efficient framework and higher benefits for data shuffling.

Data shuffling is used in computational tasks, such as large-scale distributed machine learning over massive data, where local data needs to be shuffled over iterations to train a more robust

L. Song and C. Fragouli are with the Department of Electrical and Computer Engineering, University of California, Los Angeles. T. Zhao is with the Department of Electronic Engineering, Tsinghua University. Email: {songlinqi, christina.fragouli}@ucla.edu, zhaotc13@mails.tsinghua.edu.cn. This paper was presented in part at *2017 IEEE International Symposium on Information Theory (ISIT 2017)* [1].

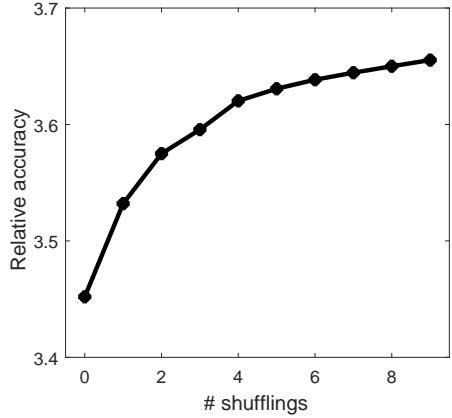

Fig. 1: An illustration of data shuffling gain over the CIFAR-10 dataset [10]. The training data size is set to be 5000 and the trained model is run over entire dataset of size 60000 to get the classification accuracy. The deep neural network is set to be 2 intermediate layers with 50 and 20 nodes on the two layers. Experiments are carried out 100 times and the figure shows the average performance in terms of the relative accuracy over random classification. We can see that after 9 times of data shuffling, the accuracy increases 6% (data shuffling gain).

model or and to achieve sufficient statistical performance [6], [7], [8], [9]. We show a small example we run in Fig. 1, where we use a deep neural network to train a classifier, and show how the performance of the classifier improves as a function of the data shuffling.

We focus on the "master-workers" distributed computing model [2], [3], where a mester node has $m$ messages and is connected through a broadcast channel to $n$ worker nodes. Each worker $i$ is equipped with a cache that can store $s_i$ messages. The computation occurs in interations: in each iteration worker nodes carry out local computations based on their cache data, and output local outcomes; the master node aggregates these local outcomes to obtain a global outcome; then the master performs data shuffling by sending new messages to refresh the cache of each node $i$. Application examples include distributed machine learning in datacnters, where data shuffling updates the training data in workers [2], and mobile cloud gaming systems where each iteration equips the users with new attributes, e.g., new maps [11].

The insightful idea of using index coding in data shuffling is as follows [12], [2]. At each iteration, the master node randomly interleaves the $m$ messages, and then allocates some specific $s_i$ messages to each worker $i$. This is equivalent to an index coding problem, where each worker makes some specific $s_i$ requests, and messages that worker nodes have from previous iterations form side information. Index coding aims to find the smallest amount of broadcast transmissions to satisfy all requests, and can offer benefits over uncoded broadcast transmissions. However, index coding has been shown to be NP-hard and in the worst case may require $\Omega(n)$ transmissions [12]. For random graph instances, it almost surely requires $\Theta(n/\log(n))$ transmissions [13], [14].

The main observation behind our work is that, when performing data interleaving, there exist multiple choices for the interleaver that can lead to an equally good performance. For example, to train a classification model in a distributed system, large volume of data instances can be randomly distributed into $n$ worker nodes in tens of millions of ways. Generally, the number of possible permutations increase exponentially in the number of workers $n$ and the number of messages cached in each worker $s_i$, while the number of iterations is often no more than polynomial of $n$ and $s_i$. Thus, we can consider a pliable index coding framework, and jointly decide the communication coding scheme and the messages to send to each worker, so that the number of broadcast transmissions is minimized. Pliable index coding, introduced in [15], considers, like index coding, a server that has $m$ messages connected through a broadcast channel to $n$ clients; however, now the clients are happy to receive any message they do not already have. This degree of freedom enable us to design more efficient coding and transmission schemes: pliable index coding requires in the worst case $O(\log^2(n))$, an exponentially smaller number of transmissions than index coding, and these benefits can be achieved in polynomial time [15], [16].

Another way of phrasing our observation is that, when performing data shuffling, if instead of completely random data interleaving, we perform semi-random interleaving, we can design the semi-random interleaver jointly with the coding scheme, so that the communication cost is significantly reduced, while still achieving "good" shuffling of the data. To measure how "good" the data shuffling is, we introduce an average Hamming distance metric, that captures the fact that we want the cached content to be different across workers and iterations [2], [17], [9], [18].

The main contribution of this paper is the design of a semi-randomized data shuffling and coding scheme for distributed computing, that ensures a desired level of average Hamming distance, and builds on two (significant) modifications of the pliable index coding design, described next.

The first modification aims to reduce the correlation of cached content across workers: when conducting data shuffling, we want each message to go to at most a specific number of workers, say $c$, so as to achieve an unbiased data distribution that looks "random-like". We capture this by imposing the constraint that each message can be used to satisfy at most $c$ clients. That is, each client is happy to receive any message she does not have, but at most $c$ clients can receive the same message. We show that even if $c = 1$, i.e., each message can satisfy at most one client, we can still achieve $O(n)$ benefits over index coding in some cases; this is because, we still have the freedom to select any of the $O(n!)$ interleaved versions of requests that lead to the smallest number of transmissions. We prove that the constrained pliable index coding problem is NP-hard. We show that for random instances, the optimal code length is almost surely upper bounded by $O(\min\{\frac{n}{c\log(n)}, \frac{n}{\log(m)}\})$ for $c = o(\frac{n^{1/7}}{\log^2(n)})$ and $O(\min\{\frac{n}{c} + \log(c), \frac{n}{\log(m)}\})$ for $c = \Omega(\frac{n^{1/7}}{\log^2(n)})$. We believe that our results in constrained pliable index coding are theoretically interesting in their own, beyond the particular application to data shuffling.

The second modification aims to reduce the correlation of cached content across iterations: we design a hierarchical transmission scheme for data shuffling that utilizes constrained pliable index coding as a component. We divide the messages and workers into groups and for each group we perform a constrained pliable index coding to shuffle the messages. We show that our scheme can achieve benefits $O(ns/m)$, in terms of transmissions over index coding, with linear encoding complexity at the master node, where $s$ is the cache size and $ns/m$ is the average number of workers that cache each message.

We experimentally evaluate of data shuffling scheme over a real dataset in a distributed classification problem. The results show that our proposed pliable index coding based semi-random shuffling scheme on average can save $88\%$ of the transmissions, with a $2\%$ performance loss in terms of error rate, compared with the index coding based random shuffling scheme.

The paper is organized as follows. We describe our model in Section II. We present our formulation and results in constrained pliable index coding in Section III. We introduce our hierarchical structure for data shuffling, that uses constrained pliable index coding as a component, in Section IV. We provide experimental results in Section V and conclude the paper in Section VI.

## II. Model and Metrics

### A. Distributed Computing System Model

We consider a "master-workers" distributed computing system, with one master node that has $m$ messages $b_1, b_2, \ldots, b_m$ in a finite field $\mathbb{F}_q$ and $n$ worker nodes (or clients in the pliable index coding framework). Throughout the paper, we will use $[y]$ ($y$ is a positive integer) to denote the set $\{1, 2, \ldots, y\}$ and use $|Y|$ to denote the cardinality of set $Y$. We will interchangeably use $b_j$ and message $j \in [m]$ to refer to messages and similarly $c_i$ or $i \in [n]$ to refer to workers (clients). Each worker $i \in [n]$ is equipped with a cache of size $s_i$. The master node can make error-free broadcasting transmissions to all workers.

The system aims at solving a computational task $y = f(b_1, b_2, \ldots, b_m)$, where $f$ is a function of all messages (e.g., data instances in distributed machine learning). Distributed computing achieves this task through iterations. Initially, each worker $i \in [n]$ has cached some subset of the messages indexed by $S_i \subseteq [m]$ and an initial value of the outcome $y^0$ is broadcasted to all workers. The following procedures are operated for each iteration $t = 1, 2, \ldots$:

1) Each worker $i$ performs local computation $y_i^t = f_i(\{b_j\}_{j \in S_i}, y^{t-1})$, where $f_i$ is a function of local messages cached at worker $i$ and the initial value $y^{t-1}$.

2) Each worker $i$ returns the local outcome $y_i^t$ to the master node. The master node combines all the local outcomes to get a global outcome $y^t$ and broadcasts to all workers as the initial value for next iteration.

3) The system performs data shuffling: the master node makes broadcast transmissions (that may be encoded) to all workers and each worker replaces some of the old messages with new.

### B. Performance Metric for Data Shuffling

Motivated by the fact that a "good" data shuffling needs the cached content be sufficiently different across workers and iterations [2], [17], [9], [18], we use an average Hamming distance metric to measure the effect of semi-random data shuffling based on the difference of cached messages across workers and iterations. We define the cache state of worker node $i$ at iteration $t$ to be an indicator $z_i^t \in \{0, 1\}^m$, where the $j$-th bit of $z_i^t$, denoted by $z_i^t(j)$, takes value 1 if message $b_j$ is in the cache of worker $i$ at the beginning of iteration $t$ and 0 otherwise. The Hamming distance between two indicators $z$ and $z'$, denoted by $H(z, z')$, is the number of positions where the entries are different for $z$ and $z'$. We define the Hamming distance of a shuffling scheme as the average Hamming distance across time and worker nodes $H \triangleq \frac{1}{\binom{Tn}{2}} \sum_{t,t' \in [T], i,i' \in [n], (t,i) \neq (t',i')} \mathbb{E}[H(z_i^t, z_{i'}^{t'})]$, where $T$ denotes the number of iterations. We note that the average Hamming distance achieved by uniform at random data shuffling is $s(1 - \frac{s}{m})$. Note that there are also some other metrics to measure the data shuffling performance, such as the Kullback-Leibler divergence, the Wasserstein distance, however, as a first step research, Hamming distance metric is the most straightforward one and other metrics can be converted to Hamming distance in the relevant domains.

## III. Constrained Pliable Index Coding

We here present our formulation and results for constrained pliable index coding, that forms a core component of the hierarchical data shuffling scheme described in the next section.

### A. Problem Formulation

We consider a server with $m$ messages $b_1, b_2, \ldots, b_m$ in a finite field $\mathbb{F}_q$ connected through a lossless broadcast domain to $n$ clients. Each client $i \in [n]$ has as side information some subset of the messages, indexed by $S_i \subseteq [m]$, and is happy to receive any one of the remaining messages, indexed by $R_i = [m] \backslash S_i$. We term the set $R_i$ the request set. We would like to minimize the number of broadcast transmissions required to satisfy all clients, under a *c-constraint*: we require that each message $j$ is decoded and cached by at most $c$ clients who request this message. We call such a problem $c$-constrained pliable index coding and denote it by $(m, n, \{R_i\}_{i \in [n]}, c)$. In this work, we focus on scalar linear coding as we describe next; we note that vector linear coding was shown to not offer order-of-magnitude benefits for pliable index coding [16]: both scalar and vector linear pliable index coding achieve the lower bound $\Omega(\log(n))$ and the upper bound $O(\log^2(n))$ for the optimal number of broadcast transmissions.

*Linear Encoding:* The server makes $L$ broadcast transmissions $x_1, x_2, \ldots, x_L$ over a noiseless channel. Each $x_l$ is a linear combination of the messages $b_1, \ldots, b_m$, namely, $x_l = a_{l1}b_1 + a_{l2}b_2 + \ldots + a_{lm}b_m$, where $a_{lj} \in \mathbb{F}_q$ are the coding coefficients. We refer to the number of transmissions, $L$, as the *code length* and to the $L \times m$ matrix $\boldsymbol{A}$ with entries $a_{lj}$ as the *coding matrix*. In matrix form, we can write

$$\boldsymbol{x} = \boldsymbol{A}\boldsymbol{b}, \tag{1}$$

where $\boldsymbol{b}$ and $\boldsymbol{x}$ are vectors that collect the original messages and coded transmissions, respectively.

*Linear Decoding:* Given $\boldsymbol{A}$, $\boldsymbol{x}$, and $\{b_j | j \in S_i\}$, each client $i$ needs to solve the linear equation (1) to get a unique solution of $b_{j_i}$, for some $j_i \in R_i$. We say that client $i$ is satisfied if he/she stores the decoded message $b_{j_i}$ and $b_{j_i}$ is decoded and stored by at most $c$ clients. Clearly, client $i$ can remove from the transmissions his/her side information messages, i.e., to recover $x_l^{(i)} = x_l - \sum_{j \in S_i} a_{lj}b_j$ from the $l$-th transmission. As a result, client $i$ only needs to solve

$$\boldsymbol{A}_{R_i}\boldsymbol{b}_{R_i} = \boldsymbol{x}^{(i)}, \tag{2}$$

to retrieve a message $b_{j_i}$ she does not have, where $\boldsymbol{A}_{R_i}$ is the sub-matrix of $\boldsymbol{A}$ with columns indexed by $R_i$; $\boldsymbol{b}_{R_i}$ is the message vector with elements indexed by $R_i$; and $\boldsymbol{x}^{(i)}$ is a $L$-dimensional column vector with elements $x_l^{(i)}$.

The following decoding criterion was derived in [16] and repeated in the context of c-constrained coding here. We use $\boldsymbol{a}_j$ to denote the $j$-th column of matrix $\boldsymbol{A}$ and use $\text{span}\{\boldsymbol{a}_{j'}|j' \in R_i \backslash \{j\}\} = \{\sum_{j' \in R_i \backslash \{j\}} \lambda_{j'} \boldsymbol{a}_{j'}|\lambda_{j'} \in \mathbf{F}_q\}$ to denote the linear space spanned by columns of $\boldsymbol{A}$ indexed by $R_i$ other than $j$.

**Lemma 1.** *In a constrained pliable index coding problem* $(m, n, \{R_i\}_{i \in [n]}, c)$*, a coding matrix* $\boldsymbol{A}$ *can satisfy all clients if and only if there exist messages* $j_1, j_2, \ldots, j_n \in [m]$*, one for each client, where no single message is repeated more than* $c$ *times, i.e.,* $j_{i_1} = j_{i_2} = \ldots = j_{i_{c+1}}$ *does not hold for any combination of* $c+1$ *clients* $i_1, i_2, \ldots, i_{c+1} \in [n]$*, such that the matrix* $\boldsymbol{A}$ *satisfies*

$$\boldsymbol{a}_{j_i} \notin \text{span}\{\boldsymbol{a}_{j'}|j' \in R_i \backslash \{j\}\}, \forall i \in [n]. \tag{3}$$

*Bipartite Graph Representation:* We sometimes use a bipartite graph representation of the pliable index coding problem, where on one side the vertices correspond to messages and on the other side to clients; we connect clients to the messages they *do not* have, i.e., client $i$ connects to the messages in $R_i$ [19].

*Design Goal:* Our goal is to construct a coding matrix $\boldsymbol{A}$ that satisfies all clients with the minimum code length $L$. Note that the $c$-constraint significantly changes the pliable index coding problem. For example, assume we have $m$ messages and $n$ clients with no side information; then pliable index coding requires 1 transmission, while constrained pliable index coding needs $n/c$ transmissions to satisfy all clients.

## B. Benefits Over Index Coding

Clearly, the larger the value of $c$, the more benefits we expect constrained pliable index coding to have over index coding (for $c = n$ we have exponential benefits [15], [16]). We here provide an example to show that it is possible to have benefits of $O(n)$ even when $c = 1$, i.e., each message can satisfy at most one client, as is the case in index coding. This equivalently shows that, if we are allowed to "interleave the demands" in index coding, we can gain $O(n)$ in terms of the number of transmissions.

We construct the following 1-constrained pliable coding instance with $n$ messages and $n$ clients. Client $i \in [n/2]$ requests any of the messages 1 to $n/2$ and $n/2 + i$, i.e., $R_i = \{1, 2, \ldots, n/2, n/2 + i\}$, for $i \in [n/2]$. Client $i \in [n] \backslash [n/2]$ requests any of the messages $n/2 + 1$ to $n$ and $i - n/2$, i.e., $R_i = \{i - n/2, n/2 + 1, n/2 + 2, \ldots, n\}$, for $i \in [n] \backslash [n/2]$. All messages not in the request set form side information.

For index coding, if client $i$ requests message $i$ and has the same side information as above, then we need at least $n/2$ transmissions, since the first $n/2$ clients do not have the first $n/2$ messages as side information. In contrast, 1-constrained pliable index coding only requires 2 transmissions. Indeed, we can enable client $i \in [n/2]$ to decode the message $n/2 + i$, by making the transmission $b_{n/2+1} + b_{n/2+2} + \ldots + b_n$, since each client $i \in [n/2]$ has all messages indexed by $[n] \backslash ([n/2] \cup \{n/2 + i\})$ as her side information. Similarly, we can enable client $i \in [n] \backslash [n/2]$ to decode the message $i - n/2$ by making the transmission $b_1 + b_2 + \ldots + b_{n/2}$.

## C. Constrained Pliable Index Coding is NP-hard

It suffices to show that 1-constrained pliable index coding is NP-hard.

**Theorem 1.** *For a 1-constrained pliable index coding problem, deciding if the optimal code length*

- $L = 1$ *is in P.*
- $L = 2$ *is NP-complete.*

The $L = 1$ case is easy to see: if one transmission can make each client to receive a distinct message, then the server needs to linearly combine exactly $n$ messages, one for each client. Client $i$ can decode a message $b_j$, $j \in R_i$, only if all other $n - 1$ messages are in her side information set. A greedy approach enables to test whether such $n$ messages exist can be tested in polynomial time. For $L = 2$, we use a reduction from the graph coloring problem, see Appendix A for the complete theorem proof.

## D. Lower and Upper Bounds for Constrained Pliable Index Coding

In this subsection, similar to index coding, we show that there is also a "sandwich property" for the constrained pliable index coding problem, bounding the optimal code length from above and below.

*A Lower Bound:* For the conventional index coding problem, a lower bound on the number of required transmissions equals the maximum independent set of an undirected graph with $n$ vertices and an edge connecting two vertices $i$ and $i'$ if and only if clients $i$ and $i'$ do not have messages $i'$ and $i$ as their side-information [12]. We can interpret this bound as follows: assume $k$ is the size of the largest set of clients (and their corresponding required messages), such that no one of them has any of the corresponding required messages as side-information. Then the server needs to make at least $k$ broadcast transmissions, to convey the $k$ required messages to the $k$ clients, since none of these clients has the other clients required messages as side information.

For the $c$-constrained pliable index coding scenario, the index coding lower bound does not hold since the same message can satisfy $c$ clients, e.g., for the instance in Section III-B. We derive a lower bound that follows the same spirit, identifying group of clients that require a certain number of transmissions to be satisfied, but that uses a different approach based on our decoding criterion, as described in the next theorem.

**Theorem 2.** *In a c-constrained pliable index coding instance, if there exist $k$ clients $i_1, i_2, \ldots, i_k$, such that the request sets satisfy $R_{i_1} \subseteq R_{i_2} \subseteq \ldots \subseteq R_{i_k}$, then the code length is at least $k/c$.*

*Proof.* From the decoding criterion, we assume that there exist messages $j_1 \in R_{i_1}, j_2 \in R_{i_2}, \ldots, j_k \in R_{i_k}$, such that for the coding matrix $\boldsymbol{A}$, $\boldsymbol{a}_{j_s} \notin \text{span}\{\boldsymbol{A}_{R_{i_s} \setminus \{j_s\}}\}$ for $s = 1, 2, \ldots, k$. Since $R_{i_1} \subseteq R_{i_2} \subseteq \ldots \subseteq R_{i_k}$, we have $\boldsymbol{a}_{j_k} \notin \text{span}\{\boldsymbol{a}_{j_1}, \boldsymbol{a}_{j_2}, \ldots, \boldsymbol{a}_{j_{k-1}}\}$, $\boldsymbol{a}_{j_{k-1}} \notin \text{span}\{\boldsymbol{a}_{j_1}, \boldsymbol{a}_{j_2}, \ldots, \boldsymbol{a}_{j_{k-2}}\}$, $\ldots$, $\boldsymbol{a}_{j_2} \notin \text{span}\{\boldsymbol{a}_{j_1}\}$, which implies that after removing redundancy, the set of vectors $\{\boldsymbol{a}_{j_1}, \boldsymbol{a}_{j_2}, \ldots, \boldsymbol{a}_{j_k}\}$ are linearly independent. Hence, the coding matrix $\boldsymbol{A}$ needs to be have a rank at least $k/c$ and the result follows. $\qquad\square$

*An Upper Bound:* To derive our upper bound, we find the smallest number $k$ of colors to color the vertices of bipartite graph that represents a problem instance, so that the coloring scheme satisfies the following two properties:

- Each client vertex $i \in [n]$, has exactly 1 neighbor that has the same color;
- Each message vertex $j \in [m]$, has at most $c$ neighbors that have the same color.

If such a coloring exists, then we can satisfy all clients with $k$ transmissions. Each of the $k$ transmissions consists of a linear combination (with coefficient 1) of messages that have the same color, i.e., $\sum b_{j'}$ for all $b_{j'}$ with the same color. It is not hard to see that this transmission scheme can result in successful coding. Indeed, for client $i$, if the color is 'red', then the transmission corresponding to the color 'red' can help client $i$ recover the only neighbor of $i$ that has the same color 'red'; and also any message vertex with a color 'red' will be recovered and stored by at most $c$ neighbors with the same color 'red'.

We can see that this minimum number of colors is just the *partition number* $\mathcal{P}_{star}(G)$ of the graph $G$, where $\mathcal{P}_{star}(G)$ is the minimum number of *induced star forests*[1] into which the graph can be partitioned such that any induced star is centered on a message vertex in $[m]$ with degree no more than $c$. An example is shown in Fig. 2. Note that as a special case when $c = 1$, this partition number is the minimum number of *induced matchings*[2] into which the graph can be partitioned.

### E. Performance Over Random Instances

We consider a random bipartite graph instance, denoted by $B(m, n, p)$, or $B$ for short, where there are $m$ messages and $n$ clients, each message can be recovered and cached by at most $c$ clients, and each client is connected with a message with probability $p$ (recall that clients have as side information all the messages they are not connected to). We assume that $p$ is a fixed constant and define $\bar{p} = \min\{p, 1 - p\}$, while $c = c(n)$ and $m = m(n) \geq n$ could be changing with $n$. Hence, in the following, $o(1)$ refers to $\lim f(n) = 0$ as $n \to \infty$.

Theorem 3 summarizes our main result. We then provide a proof outline, followed by a complete proof.

**Theorem 3.** *The number of broadcast transmissions for random graph instance $B(m, n, p)$ with $c$-constraint is almost surely upper bounded by*

- $O(\min\{\frac{n}{c \log(n)}, \frac{n}{\log(m)}\})$, *for* $c = o(\frac{n^{1/7}}{\log^2(n)})$; *and*
- $O(\min\{\frac{n}{c} + \log(c), \frac{n}{\log(m)}\})$, *for* $c = \Omega(\frac{n^{1/7}}{\log^2(n)})$.

Our proof outline is as follows (except a simple scenario that we will discuss later). We design a transmission scheme, and show that it achieves this performance. To do so, we first define a $k$-*pattern* to be an induced star forest (we will give details later) that enables with a single broadcast transmission to satisfy $kc$ clients. We then find values $k = \mathbb{K}$, $m'$ and $n'$ for which almost surely a $\mathbb{K}$-pattern exists in every induced subgraph $B'$ of $B$ with $m'$ message vertices and $n'$ client vertices. Let us denote by $\mathcal{B}(B, m', n')$ the family of all induced subgraphs of $B$ by $m'$ message vertices and $n'$ client vertices. Then this condition can be expressed formally as

$$\Pr\{\exists B' \in \mathcal{B}(B, m', n'), s.t., B' \text{ contains no } \mathbb{K}\text{-patterns}\} = o(1). \tag{4}$$

---

[1]A star is a complete bipartite graph $K_{1,l}$ with degree $l$. An induced star forest is an induced subgraph consists of disjoint stars. An induced subgraph is a subset of the vertices of a graph together with any edges whose endpoints are both in this subset of vertices.

[2]A matching in a graph is an induced matching if it occurs as an induced subgraph of the graph.

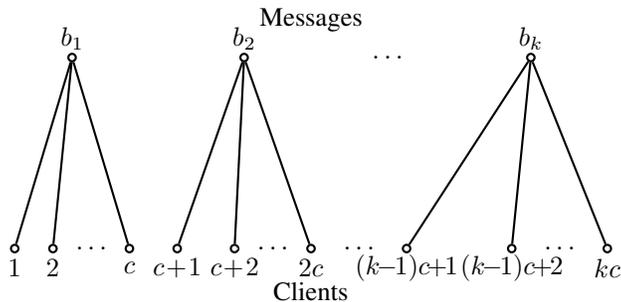

Fig. 2: In a $k$-pattern, one transmission satisfies $kc$ clients.

The transmission scheme, termed *RandTrans*, proceeds as follows. If there are more than $n'$ clients in the original graph $B$, we pick a $\mathbb{K}$-pattern and make one transmission. We remove the satisfied clients and the used messages. We repeat this again and again until there are less than $n'$ clients. We then use at most $n'$ transmissions to satisfy the remaining clients. Hence, we almost surely need $\frac{n}{\mathbb{K}c} + n'$ transmissions.

To minimize $\frac{n}{\mathbb{K}c} + n'$, we want $n'$ to be small and $\mathbb{K}$ to be large. However, by decreasing $n'$ we also decrease the values of $\mathbb{K}$ that satisfy (4). Hence, we need to balance the sizes of $n'$ and $\mathbb{K}$; we use different values of $\mathbb{K}$ depending on how $m$, $n$, and $c$ are related.

Next, we provide the complete proof. First, we define an induced subgraph called *$k$-pattern*.

**Definition 1.** *A $k$-pattern is an induced subgraph that consists of $k$ message vertices and $kc$ client vertices, where each of the $k$ messages is connected with $c$ distinct clients and each of the $kc$ client is connected with only one message.*

Essentially, a $k$-pattern is an induced star forest, as illustrated in Fig. 2. For a given random bipartite graph $B$, let us denote by $Y_k$ the number of $k$-patterns in $B$. For an induced subgraph $B'$ of $B$, let us denote by $Y_k^{B'}$ the number of $k$-patterns contained on the subgraph $B'$. We can calculate the average number of $k$-patterns as follows:

$$\mathbb{E}[Y_k] = \binom{m}{k}\binom{n}{kc}\binom{kc}{c, c, \ldots, c}p^{kc}(1-p)^{k(k-1)c}, \tag{5}$$

where $\binom{kc}{c,c,\ldots,c} = \frac{(kc)!}{(c!)^k}$ denotes the multinomial coefficient.

It is easy to see that $\mathbb{E}[Y_k]$ is decreasing with $k$, given other parameters fixed. Hence, we define $k_0$ to be the maximum integer such that $\mathbb{E}[Y_{k_0}] \geq 1$, i.e., $k_0 = \max\{k|\mathbb{E}[Y_k] \geq 1\}$. We next show that $k_0$ is in the order of $\log(n) + \frac{\log(m)}{c} - \log(c)$. More accurately, we prove the following lemma.

**Lemma 2.** $k_0$ *satisfies:* $x_1 \leq k_0 \leq x_1 + O(1)$ *for* $x_1 = 1 + \frac{1}{\log(1/(1-p))}[\log(n) + \frac{\log(m)}{c} - \log(c) - \frac{\log(\log(n)+\frac{\log(m)}{c}-\log(c))}{c} - \frac{\log\log(\frac{1}{1-p})}{c} + \log(p)] + o(1).$

*Proof.* Using the binomial inequality

$$(\frac{x}{y})^y \leq \binom{x}{y} \leq (\frac{ex}{y})^y, \tag{6}$$

we can bound the value of $\mathbb{E}[Y_k]$ by

$$\begin{aligned}
\mathbb{E}[Y_k] &\leq (\frac{em}{k})^k(\frac{en}{c})^c(\frac{e(n-c)}{c})^c(\frac{e(n-2c)}{c})^c \cdots (\frac{e(n-kc)}{c})^c p^{kc}(1-p)^{k(k-1)c} \\
&< (\frac{em}{k})^k(\frac{en}{c})^{kc}p^{kc}(1-p)^{k(k-1)c},
\end{aligned} \tag{7}$$

and

$$\begin{aligned}
\mathbb{E}[Y_k] &\geq (\frac{m}{k})^k(\frac{n}{c})^c(\frac{n-c}{c})^c(\frac{n-2c}{c})^c \cdots (\frac{n-kc}{c})^c p^{kc}(1-p)^{k(k-1)c} \\
&> (\frac{m}{k})^k(\frac{n-kc}{c})^{kc}p^{kc}(1-p)^{k(k-1)c}.
\end{aligned} \tag{8}$$

By taking $\log(\cdot)$ on both sides, we get the following relationships:

$$\begin{aligned}
\log(\mathbb{E}[Y_k]) &< k[1 + \log(m) - \log(k)] + kc[1 + \log(n) - \log(c)] + kc\log(p) \\
&\quad + k(k-1)c\log(1-p), \\
\log(\mathbb{E}[Y_k]) &> k[\log(m) - \log(k)] + kc[\log(n-kc) - \log(c)] + kc\log(p) \\
&\quad + k(k-1)c\log(1-p).
\end{aligned} \tag{9}$$

Let us define two continuous functions $f_1(x) = x[\log(m) - \log(x)] + xc[\log(n - xc) - \log(c)] + xc\log(p) + x(x-1)c\log(1-p)$ and $f_2(x) = x[1 + \log(m) - \log(x)] + xc[1 + \log(n) - \log(c)] + xc\log(p) + x(x-1)c\log(1-p)$. Hence, we can rewrite the above inequalities as $f_1(k) < \log(\mathbb{E}[Y_k]) < f_2(k)$. Note that the two functions $f_1(x)$ and $f_2(x)$ are monotonously decreasing around $\log(n) + \frac{\log(m)}{c} - \log(c)$.

Recall that we are interested in the maximum integer $k_0$ such that $\log(\mathbb{E}[Y_{k_0}]) \geq 0$. Solving the equations $f_1(x) = 0$ and $f_2(x) = 0$ by iterative manipulations, we get

$$
\begin{aligned}
f_1(x_1) = 0, \text{ for } x_1 = &\ 1 + \tfrac{1}{\log(1/(1-p))}[\log(n) + \tfrac{\log(m)}{c} - \log(c) \\
&- \tfrac{\log(\log(n) + \frac{\log(m)}{c} - \log(c))}{c} - \tfrac{\log\log(\frac{1}{1-p})}{t} + \log(p)] + o(1), \\
f_2(x_2) = 0, \text{ for } x_2 = &\ 1 + \tfrac{1}{\log(1/(1-p))}[\log(n) + \tfrac{\log(m)}{c} - \log(c) + 1 + \tfrac{1}{c} \\
&- \tfrac{\log(\log(n) + \frac{\log(m)}{c} - \log(c))}{c} - \tfrac{\log\log(\frac{1}{1-p})}{c} + \log(p)] + o(1).
\end{aligned}
\tag{10}
$$

We can see that both $x_1$ and $x_2$ are in the order of $\log(n) + \frac{\log(m)}{c} - \log(c)$ and $x_2 - x_1 = \frac{1}{\log(1/(1-p))}[1 + \frac{1}{c}] + o(1) \leq \frac{2}{\log(1/(1-p))} + o(1)$, which is bounded by $O(1)$.

We also have $\log(\mathbb{E}[Y_{\lceil x_2 \rceil}]) < f_2(\lceil x_2 \rceil) \leq f_2(x_2) = 0$ and $\log(\mathbb{E}[Y_{\lfloor x_1 \rfloor}]) > f_1(\lfloor x_1 \rfloor) \geq f_1(x_1) = 0$. This implies that $x_1 - 1 < \lfloor x_1 \rfloor \leq k_0 \leq \lceil x_2 \rceil - 1 < x_2$, from which the result follows. $\qquad\square$

What we would like to show next is that the average number of $k$-patterns $\mathbb{E}[Y_k]$ has the property that it changes fast around the value $k_0$. We have the following lemma.

**Lemma 3.** $\mathbb{E}[Y_{k_1}]$ *satisfies:* $\mathbb{E}[Y_{k_1}] \geq (\frac{n}{ec})^{3c(1+o(1))} m^{3(1+o(1))}$, *for* $k_1 = k_0 - 3$.

*Proof.* We first have the following equation

$$
\begin{aligned}
\frac{\mathbb{E}[Y_{k_0-3}]}{\mathbb{E}[Y_{k_0}]} &= \frac{\binom{m}{k_0-3}\binom{n}{(k_0-3)c}\binom{(k_0-3)c}{c}\binom{(k_0-4)c}{c}\ldots\binom{c}{c}p^{(k_0-3)c}(1-p)^{(k_0-3)(k_0-4)c}}{\binom{m}{k_0}\binom{n}{k_0 c}\binom{k_0 c}{c}\binom{(k_0-1)c}{c}\ldots\binom{c}{c}p^{k_0 c}(1-p)^{k_0(k_0-1)c}} \\
&= \frac{k_0(k_0-1)(k_0-2)(c!)^3}{(m-k_0+3)(m-k_0+2)(m-k_0+1)(n-k_0 c+1)(n-k_0 c+2)\ldots(n-k_0 c+3c)p^{3c}(1-p)^{6k_0 c - 12c}} \\
&\geq \frac{(c!)^3}{m^3 n^{3c}(1-p)^{6c(k_0-2)}} \\
&\geq (\tfrac{n}{ec})^{3c(1+o(1))} m^{3(1+o(1))},
\end{aligned}
\tag{11}
$$

where the last inequality follows from $c! \geq e(c/e)^c$ and $(1-p)^{6c(k_0-2)} = (\frac{nm^{1/c}}{c})^{6c(1+o(1))}$, since $k_0 - 2 = \frac{1}{\log(1/(1-p))}[\log(n) + \frac{\log(m)}{c} - \log(c) + o(\log(n) + \frac{\log(m)}{c} - \log(c))]$.

Also note that $\mathbb{E}[Y_{k_0}] \geq 1$ and the result follows from (11). $\qquad\square$

Similarly, we can define $k_0^{B'}$ as the maximum integer such that $\mathbb{E}[Y_{k_0^{B'}}^{B'}] \geq 1$ and define $k_1^{B'} = k_0^{B'} - 3$.

Next, we will discuss in different scenarios (in the following scenarios 1, 2, and 3) that we can find another integer $\mathbb{K} = \mathbb{K}(m,n) = k_2^{B'} \leq k_1^{B'}$, such that every induced subgraph $B' \in \mathcal{B}(B, m', n')$ almost surely contains a $\mathbb{K}$-pattern. For the fourth scenario, we will discuss separately. The 4 scenarios with parameters $\mathbb{K}$, $m'$, and $n'$ are formally defined as follows.

**Definition 2.** *We define the following 4 scenarios and how the corresponding parameters are related.*

- *Scenario 1:* $m < \exp(n^{1/15})$. *In this scenario, we set* $c = 1$, $\mathbb{K} = \lfloor \frac{1}{\log(1/\bar{p})} \lceil \log(m) - 3 \log \log(m) + 2 \log \log(\frac{1}{\bar{p}}) \rceil \rfloor = \Theta(\log(m))$, $m' = \frac{m}{\log(m)}$, *and* $n' = \frac{n}{\log(m)}$. *If* $c > 1$, *we simply set* $c = 1$ *and this is a stronger constraint.*

- *Scenario 2:* $m \geq \exp(n^{1/15})$. *In this scenario, we set* $c = 1$, $\mathbb{K} = \lfloor \frac{1}{\log(1/\bar{p})} \lceil \log(m) - 3 \log \log(m) + 2 \log \log(\frac{1}{\bar{p}}) \rceil \rfloor = \Theta(\log(m))$, $m' = m - \mathbb{K} = m(1 - o(1))$, *and* $n' = \mathbb{K}$. *If* $c > 1$, *we simply set* $c = 1$ *and this is a stronger constraint.*

- *Scenario 3:* $c = o(\frac{n^{1/7}}{\log^2(n)})$. *In this scenario, we set* $\mathbb{K} = \lfloor \frac{1}{\log(1/\bar{p})} \lceil \log(n) - 3 \log \log(n) - 3 \log(c) + 2 \log \log(\frac{1}{\bar{p}}) \rceil \rfloor = \Theta(\log(n))$, $m' = \frac{m}{\log(n)}$, *and* $n' = \frac{n}{c \log(n)}$.

- *Scenario 4:* $c = \Omega(\frac{n^{1/7}}{\log^2(n)})$. *In this scenario, we set* $\mathbb{K} = 1$, $m' = 1$, *and* $n' = \frac{2c}{p}$.

Note that the scenarios 1 and 2 are defined based on the relationship between $m$ and $n$; the scenarios 3 and 4 are defined based on the relationship between $c$ and $n$. There maybe overlaps between scenarios $1, 2$ and scenarios $3, 4$. We want to show the following lemma for the first 3 scenarios.

**Lemma 4.** *For scenarios* 1, 2, *and* 3 *with parameters defined in Definition 2, every induced subgraph* $B' \in \mathcal{B}(B, m', n')$ *almost surely contains a* $\mathbb{K}$-*pattern:*

$$\Pr\{\exists B' \in \mathcal{B}(B, m', n'), s.t., B' \text{ contains no } \mathbb{K}\text{-pattern}\} = o(1). \tag{12}$$

Lemma 4 is the key step to prove Theorem 3. The idea is to show that the number of $\mathbb{K}$-patterns in every $B'$ is closely concentrated around its expected number by Azuma's inequality. We will give the proof of scenario 1 in the following and the detailed proof of scenarios 2 and 3 is given in Appendix B.

*Proof.* To prove this lemma, we first use an "edge exposure" process to form a martingale based on the random subgraph $B'$ [20], [21]. Specifically, we define $X$ as a maximum number of $\mathbb{K}$-patterns in $B'$ such that no two of them share a same message-client pair (i.e., any two $\mathbb{K}$-patterns either have no common message vertices or client vertices or both). We label the possible edges as $1, 2, \ldots, m'n'$ and denote by $Z_l$ the random variable to indicate whether the edge $l$ is exposed in the random graph, i.e., $Z_l = 1$ if the $l$-th possible edge is present in the graph and $Z_l = 0$ otherwise. Therefore, $X = f(Z_1, Z_2, \ldots, Z_{m'n'})$ is a function of the variables $Z_l$. Define $X_l = \mathbb{E}[X|Z_1, Z_2, \ldots, Z_l]$ as a sequence of random variables for $l = 1, 2, \ldots, m'n'$, then $\{X_l\}$ is a Doob martingale and $X_{m'n'} = X$. Obviously, the function $X = f(Z_1, Z_2, \ldots, Z_{m'n'})$ is 1-Lipschitz, namely, flipping only one indicator function, some $Z_l$, the value of $X$ differs by at most 1: $|f(Z_1, \ldots, Z_l, \ldots, Z_{m'n'}) - f(Z_1, \ldots, Z_{l-1}, 1 - Z_l, Z_{l+1}, \ldots, Z_{m'n'})| \leq 1$.

Note that the subgraph $B'$ contains no $\mathbb{K}$-pattern is equivalent to $X = 0$. We then use the Azuma's inequality

$$\Pr\{\mathbb{E}[X] - X \geq a\} \leq \exp(-\frac{a^2}{2m'n'}), \text{ for } a > 0. \tag{13}$$

to get

$$\Pr\{X = 0\} = \Pr\{\mathbb{E}[X] - X \geq \mathbb{E}[X]\} \leq \exp(-\frac{\mathbb{E}^2[X]}{2m'n'}). \tag{14}$$

Hence, to bound $\Pr\{X = 0\}$, we only need to find a lower bound of $\mathbb{E}[X]$. We use the following probabilistic argument. For subgraph $B'$, we define $\mathcal{K}$ as the family of all $\mathbb{K}$-patterns and $\mathcal{P}$ as the family of all $\mathbb{K}$-pattern pairs that share at least a same message and a same client. Let us denote by $B_1, B_2 \in \mathcal{B}(B', \mathbb{K}, \mathbb{K}c)$ induced subgraphs of $B'$ by $\mathbb{K}$ message vertices and $\mathbb{K}c$ client vertices. Let us also denote by $X_{B_1}$ and $X_{B_2}$ the variables to indicate whether the subgraphs $B_1$ and $B_2$ are $\mathbb{K}$-patterns. Let us use the notation $B_1 \sim B_2$ if two different subgraphs $B_1$ and $B_2$ share at least a same message vertex and a same client vertex. We then lower bound $\mathbb{E}[X]$ using the following scheme for scenarios $1, 2, 3$ (we will talk about how we bound $\mathbb{E}[X]$ for scenario 4 later): randomly select a subset of $\mathbb{K}$-patterns from the set $\mathcal{K}$ by picking up each $\mathbb{K}$-pattern with probability $p^\dagger$ (the value of which we will determine later); if two selected $\mathbb{K}$-patterns $B_1^\dagger$ and $B_2^\dagger$ form a pair in the set $\mathcal{P}$, then remove one of them. Then,

$$\mathbb{E}[X] \geq p^\dagger \mathbb{E}[|\mathcal{K}|] - p^{\dagger 2} \mathbb{E}[|\mathcal{P}|], \tag{15}$$

where the first term in the expression is the average number of selected $\mathbb{K}$-patterns in $\mathcal{K}$ and the second term is the average number of $\mathbb{K}$-patterns that are removed because a pair in $\mathcal{P}$ is selected with probability $p^{\dagger 2}$.

We observe that $\mathbb{E}[|\mathcal{K}|] = \mathbb{E}[Y_{\mathbb{K}}^{B'}]$ and next we calculate $\mathbb{E}[|\mathcal{P}|]$ and determine $p^\dagger$.

$$
\begin{aligned}
\mathbb{E}[|\mathcal{P}|] & = \tfrac{1}{2} \sum_{B_1 \in \mathcal{B}(B', \mathbb{K}, \mathbb{K}c)} \sum_{B_2 \in \mathcal{B}(B', \mathbb{K}, \mathbb{K}c): B_2 \sim B_1} \mathbb{E}[X_{B_1} X_{B_2}] \\
& = \tfrac{1}{2} \sum_{B_1} \sum_{B_2 : B_2 \sim B_1} \Pr\{X_{B_1} = 1\} \Pr\{X_{B_2} = 1 | X_{B_1} = 1\} \\
& = \tfrac{1}{2} \binom{m'}{\mathbb{K}} \binom{n'}{\mathbb{K}} \binom{\mathbb{K}c}{c,c,\dots,c} \Pr\{X_{B_0} = 1\} \sum_{B_2 : B_2 \sim B_0} \Pr\{X_{B_2} = 1 | X_{B_0} = 1\} \\
& = \tfrac{1}{2} \mathbb{E}[Y_{\mathbb{K}}^{B'}] \sum_{B_2 : B_2 \sim B_0} \Pr\{X_{B_2} = 1 | X_{B_0} = 1\},
\end{aligned}
\tag{16}
$$

where the second equality is from the conditional probability formula, the third equality is by symmetry of the selection of $B_1$ and we then take a fixed selection $B_0$ consisting of the first $\mathbb{K}$ messages and first $\mathbb{K}c$ clients.

Hence, we only need to calculate the term $\sum_{B_2 : B_2 \sim B_0} \Pr\{X_{B_2} = 1 | X_{B_0} = 1\}$ for different scenarios. We upper bound this term from above by enumerating all subgraph $B_2$ that has at least one common client vertex one common message vertex with $B_0$. In the following, we give a detailed proof for scenario 1 and the proof techniques are similar for scenarios 2 and 3. For completeness, we give details of proof for scenarios 2 and 3 in Appendix B.

1) For scenario 1, we have

$$
\sum_{B_2 : B_2 \sim B_0} \Pr\{X_{B_2} = 1 | X_{B_0} = 1\} \le \sum_{j=1}^{\mathbb{K}} \sum_{i=1}^{\mathbb{K}} \binom{\mathbb{K}}{j} \binom{m'-\mathbb{K}}{\mathbb{K}-j} \binom{\mathbb{K}}{i} \binom{n'-\mathbb{K}}{\mathbb{K}-i} \mathbb{K}! \frac{p^{\mathbb{K}(1-p)^{\mathbb{K}(\mathbb{K}-1)}}}{\bar{p}^{ij}}, \tag{17}
$$

where the inequality is because $\bar{p}^{ij} \le p^a (1-p)^{ij-a}$ for any non-negative integer $a \le ij$. Let us define the term inside the summation as $\Delta_{ij} \triangleq \binom{\mathbb{K}}{j} \binom{m'-\mathbb{K}}{\mathbb{K}-j} \binom{\mathbb{K}}{i} \binom{n'-\mathbb{K}}{\mathbb{K}-i} \mathbb{K}! \frac{p^{\mathbb{K}(1-p)^{\mathbb{K}(\mathbb{K}-1)}}}{\bar{p}^{ij}}$.

We can see that for $i = 1, 2, \dots, \mathbb{K}$ and $j = 1, 2, \dots, \mathbb{K} - 1$, we have

$$
\begin{aligned}
\frac{\Delta_{i,j+1}}{\Delta_{i,j}} & = \frac{(\mathbb{K}-j)^2}{(j+1)(m'-2\mathbb{K}+j+1)} \bar{p}^{-i} \\
& \le \frac{\mathbb{K}^2}{2(m'-2\mathbb{K}+2)} \bar{p}^{-\mathbb{K}} \\
& \le \frac{\mathbb{K}^2}{m'} \bar{p}^{-\mathbb{K}} \\
& \le \frac{\frac{1}{\log^2(1/\bar{p})} \log^2(m)}{m/\log(m)} \frac{m \log^2(1/\bar{p})}{\log^3(m)} \\
& \le 1.
\end{aligned}
\tag{18}
$$

This implies that for all $i = 1, 2, \dots, \mathbb{K}$ and $j = 1, 2, \dots, \mathbb{K}$, $\Delta_{i,j} \le \Delta_{i,1}$. Also note that for

$i = 1, 2, \ldots, \mathbb{K} - 1$,

$$\begin{aligned}
\frac{\Delta_{i+1,1}}{\Delta_{i,1}} &= \frac{(\mathbb{K}-i)^2}{(i+1)(n'-2\mathbb{K}+i+1)}\bar{p}^{-1} \\
&\leq \frac{\mathbb{K}^2}{n'\bar{p}} \\
&\leq \frac{\log(m)^3}{n\bar{p}\log^2(1/\bar{p})} \\
&= o(1),
\end{aligned}$$

where the last equality holds for $m < \exp(n^{1/10})$. Hence, $\Delta_{i,j} \leq \Delta_{1,1}$ for all $i = 1, 2, \ldots, \mathbb{K}$ and $j = 1, 2, \ldots, \mathbb{K}$.

For $\Delta_{1,1}$, we have the following

$$\begin{aligned}
\frac{\Delta_{1,1}}{\mathbb{E}[Y_{\mathbb{K}}^{B'}]} &= \frac{\mathbb{K}\binom{m'-\mathbb{K}}{\mathbb{K}-1}\mathbb{K}\binom{n'-\mathbb{K}}{\mathbb{K}-1}\mathbb{K}!p^{\mathbb{K}}(1-p)^{\mathbb{K}(\mathbb{K}-1)}\bar{p}}{\binom{m'}{\mathbb{K}}\binom{n'}{\mathbb{K}}\mathbb{K}!p^{\mathbb{K}}(1-p)^{\mathbb{K}(\mathbb{K}-1)}} \\
&= \frac{\mathbb{K}^4(m'-\mathbb{K})!(m'-\mathbb{K})!(n'-\mathbb{K})!(n'-\mathbb{K})!}{\bar{p}m'!m'!(m'-2\mathbb{K}+1)!n'!(n'-2\mathbb{K}+1)!} \\
&\leq \frac{\mathbb{K}^4\log^2(m)}{\bar{p}mn} \leq \frac{\log^6(m)}{\bar{p}mn\log^4(1/\bar{p})}.
\end{aligned}$$

Plugging into (16), we have

$$\mathbb{E}[|\mathcal{P}|] \leq \frac{1}{2}\mathbb{E}^2[Y_{\mathbb{K}}^{B'}]\mathbb{K}^2\frac{\log^6(m)}{\bar{p}\log^4(1/\bar{p})mn} \leq \mathbb{E}^2[Y_{\mathbb{K}}^{B'}]\frac{\log^8(m)}{2\bar{p}\log^6(1/\bar{p})mn}$$

From Lemmas 2 and 3, we have $k_0^{B'} = \frac{1}{\log(1/(1-p))}[\log(n)+\log(m)-2\log\log(m)-\log(\log(n)+\log(m)-2\log\log(m))-\log(\frac{1}{1-p})+\log(p)]+O(1)$ and $\mathbb{E}[Y_{k_0^{B'}-3}^{B'}] \geq (\frac{mn}{e\log^2(m)})^{3(1+o(1))}$. Obviously, we have $\mathbb{K} < k_1^{B'} = k_0^{B'}-3$ and $\mathbb{E}[Y_{\mathbb{K}}^{B'}] \geq (\frac{mn}{e\log^2(m)})^{3(1+o(1))}$. By setting the probability $p^\dagger = \frac{\bar{p}\log^6(\frac{1}{\bar{p}})mn}{\mathbb{E}[Y_{\mathbb{K}}^{B'}]\log^8(m)} < 1$, we can bound the average number of $X$, $\mathbb{E}[X]$, in e.q. (15), as

$$\mathbb{E}[X] \geq \frac{\bar{p}^2\log^6(\frac{1}{\bar{p}})mn}{2\log^8(m)}.$$

Plugging (22) into (14), we can bound the following probability

$$\Pr\{B' \text{ contains no } \mathbb{K}\text{-pattern}\} \leq \exp(-\frac{\bar{p}^2\log^{12}(1/\bar{p})mn}{8\log^{14}(m)})$$

Therefore, we can bound the probability that any subgraph $B'$ induced by $m'$ messages and $n'$ clients does not contain a $\mathbb{K}$-pattern:

$$\begin{aligned}
\Pr\{\exists B' \in \mathcal{B}(B, m', n'), s.t., B' \text{ contains no } \mathbb{K}\text{-pattern}\} &\leq \binom{m}{m'}\binom{n}{n'}\exp(-\frac{\bar{p}^2\log^{12}(1/\bar{p})mn}{8\log^{14}(m)}) \\
&\leq 2^{m+n}\exp(-\frac{\bar{p}^2\log^{12}(1/\bar{p})mn}{8\log^{14}(m)}) = o(1).
\end{aligned}$$

2) For scenario 2, see Appendix B. 3) For scenario 3, see Appendix B. $\square$

Given Lemma 4, we can express Theorem 3 in a slightly different way.

**Theorem 3′.** *The number of broadcast transmissions for random graph instance $B(m, n, p)$ with c-constraint is almost surely upper bounded by*

- $O(\frac{n}{\log(m)})$, *for any* $c \geq 1$;
- $O(\frac{n}{c \log(n)})$, *for* $c = o(\frac{n^{1/7}}{\log^2(n)})$;
- $O(\frac{n}{c} + \log(c))$, *for* $c = \Omega(\frac{n^{1/7}}{\log^2(n)})$.

*Proof.* For scenarios 1, 2, and 3, we can proceed using the transmission scheme RandTrans described earlier in this subsection. Note that this transmission scheme can be successfully carried out with probability 1 following from Lemma 4 and the fact that a subgraph of $B$ that contains $n'$ vertices on both sides almost surely have a perfect matching [22]. Hence, we can almost surely use the number of transmissions $\frac{n}{\mathbb{K}c} + n'$, from which the first two parts follow.

Now, let us prove the third part of the theorem for scenario 4 with $c = \Omega(\frac{n^{1/7}}{\log^2(n)})$. We use a slightly different but simple proof technique. By setting $n' = \frac{2c}{p}$, we use a 2-step transmission scheme.

$\bullet$ In the first step, we arbitrarily make $n/c$ uncoded transmissions. After each transmission, we remove up to $c$ satisfied clients as many as possible.

$\bullet$ In the second step, we divide the remaining unsatisfied clients into as few groups as possible, each with up to $c$ clients, and we use a pliable index coding scheme to satisfy each of the groups.

We want to show that we can almost surely satisfy at least $n - n'$ clients by using these $n/c$ uncoded transmissions in the first step. Hence, we can almost surely divide the remaining unsatisfied clients into at most $n'/c = 2/p$ groups and these groups almost surely take $\frac{2}{p}O(\log(c))$ broadcast transmissions [16].

For a fixed uncoded transmission, e.g., message $b_j$, we would like to show that the probability that this transmission cannot satisfy $c$ clients is exponentially small if the remaining unsatisfied clients are more than $n'$. Let us denote by $D$ the number of connections for message vertex $b_j$ to any $n'$ remaining client vertices. Then obviously, $\mathbb{E}[D] = n'p = 2c$ and the probability that the uncoded transmission of $b_j$ cannot satisfy $c$ clients (i.e., a 1-pattern exists) can be bounded by the following Chernoff bound:

$$
\begin{aligned}
\Pr\{b_j \text{ cannot satisfy } c \text{ clients}\} \quad &\le \Pr\{D \le c\} = \Pr\{D \le (1 - 1/2)\mathbb{E}[D]\} \\
&\le \exp(-\tfrac{2c(1/2)^2}{2}) = \exp(-\tfrac{c}{4}).
\end{aligned}
\tag{25}
$$

After $n/c$ uncoded transmissions, the probability that the number of remaining unsatisfied clients is more than $n'$ can be bounded as follows:

$$
\Pr\{n/c \text{ uncoded transmissions cannot satisfy } n - n' \text{ clients}\}
$$
$$
\le \tfrac{n}{c} \exp(-\tfrac{c}{4}) \le n^{6/7} \log^2(n) o(1) \exp(-\tfrac{n^{1/7}\Omega(1)}{4\log^2(n)}) = o(1).
\tag{26}
$$

Combining the two steps, we have that the number of broadcast transmissions is almost surely upper bounded by $n/c + \frac{2}{p} O(\log(c)) = O(n/c + \log(c))$ for scenario 4. □

Note that for Theorem 3′, we can combine the results and have the number of broadcast transmissions almost surely upper bounded by $O(\min\{\frac{n}{\log(m)}, \frac{n}{c\log(n)}\})$ for $c = o(\frac{n^{1/7}}{\log^2(n)})$ and $O(\min\{\frac{n}{\log(m)}, \frac{n}{c} + \log(c)\})$ for $c = \Omega(\frac{n^{1/7}}{\log^2(n)})$.

## IV. Hierarchical Data Shuffling Scheme

To perform data shuffling, one straightforward idea is to repeatedly use constrained pliable index coding for each iteration, and ensure that each worker node has a sufficient amount of cached messages replaced by new messages. However, even if we guarantee a certain Hamming distance of the cache states between two consecutive iterations, e.g., $H(z_i^t, z_i^{t+1})$, the Hamming distance between two non-consecutive iterations, $H(z_i^t, z_i^{t+2})$, may still be small.

To ensure a sufficient large Hamming distance averaged across iterations and workers, we here propose a two-layer architecture scheme for data shuffling: the outer layer divides the messages into groups and restricts each worker's cached content to messages in certain groups, and the inner layer applies constrained pliable index coding for each message group and associated workers. The intuition of using this hierarchical structure is that, each worker node receives messages from different message groups to increase the Hamming distance across iterations; while the sparsity of the outer layer structure guarantees a sufficient Hamming distance across workers. In this section, we describe this architecture, and show that for the case where we have $m$ messages and $n$ workers with equal cache size $s_i = s$, the benefits can be up to $O(sn/m)$ over index coding.

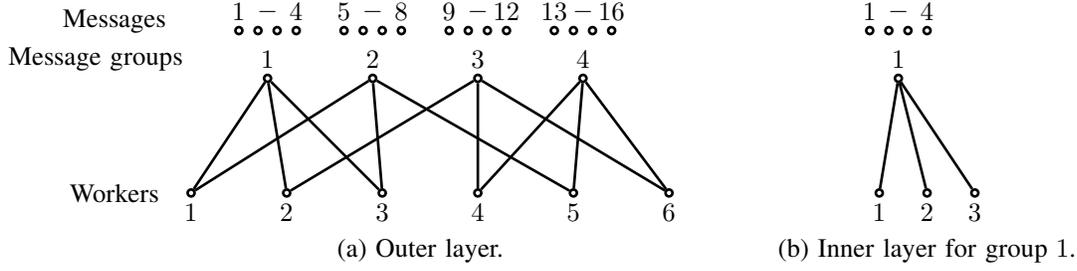

Fig. 3: An example of the two-layer scheme for data shuffling. In (a), there are $G = 4$ message groups and $n = 6$ workers. Each group has $m_1 = 4$ messages. Each worker $i$ caches messages in $|D(i)| = 2$ groups and each message group $g$ allocates messages to $|N(g)| = 3$ workers. In (b), the scheme considers the inner layer as a constrained pliable index coding instance for group 1.

### A. Hierarchical Structure and Transmission Scheme

*Outer Layer:* We partition the messages into $G = m/m_1$ groups so that every group $g \in [G]$ contains $m_1$ disjoint messages where $m_1$ is a design parameter to be decided. Let us denote by $M_1, M_2, \ldots, M_G$ the subsets of messages for groups $1, 2, \ldots, G$. In our scheme, each worker $i$ gets allocated messages from groups indexed by a set $D(i) \subseteq [G]$; each group $g$ allocates messages to workers indexed by a set $N(g) \subseteq [n]$. We can represent this relationship using a bipartite graph, as shown in Fig. 3(a): at one side there are $G$ groups, and at the other side there are $n$ workers; there is a connection between worker $i$ and group $g$ if and only if worker $i$ caches messages from group $g$, i.e., $g \in D(i)$ or $i \in N(g)$; the degree of the worker node $i$ is $|D(i)|$ and of the group node $g$ is $|N(g)|$. This structure is maintained for all iterations.

Given that we require large Hamming distance $H$, we impose the constraint that $|D(i) \cap D(i')| \leq 1$ for any two different worker pairs $i$ and $i'$, namely, they have common messages in no more than one group. Moreover, to balance the messages cached in different worker nodes, we would like that $|N(g)|$ is the same for all groups and $|D(i)|$ is the same for all workers. We thus select for our scheme to use $|D(i)| = \frac{s}{m_1(1-1/r)}$, and $|N(g)| = \frac{ns}{m_1(1-1/r)}$, where the design parameter $2 \leq r \leq m_1$ takes integer values[3]. That is, in the bipartite graph representation, all worker vertices have the same degree $|D(i)|$ and all group vertices have the same degree $|N(i)|$. We next formally define what we call a *Sparse Group Caching* structure.

**Definition 3** (**Sparse Group Caching Structure**). *The sparse group caching structure is an outer layer allocating groups of messages $g \in G$ to workers $i$, that satisfies the following conditions.*
*1)* **Equal degrees**. *Each worker is connected to $|D(i)| = \frac{s}{m_1(1-1/r)}$ groups, for all $i \in [n]$ and*

---

[3]For easy of analysis, we assume that the values $m_1(1 - 1/r)$, $|D(i)|$, and $|N(g)|$ are all integers.

*each group is connected to $|N(g)| = \frac{ns}{m(1-1/r)}$ workers, for all $g \in [G]$.*

*2)* **Balance***. At each iteration $t$, each worker $i$ caches an equal number of messages from every group $g$ in $D(i)$. Let $z_i^t(j)$ be the indicator function on whether message $j$ is cached from worker $i$ during iteration $t$, then $\sum_{j \in M_g} z_i^t(j) = m_1(1 - 1/r)$ for all $i \in [n]$, $t \in [T]$, $g \in D(i)$; and $z_i^t(j) = 0$ for all $i \in [n]$, $t \in [T]$, $j \in [m] \setminus \cup_{g \in D(i)} M_g$.*

*3)* **Sparse Connectivity.** *Any two workers may be connected to at most one common group: $|D(i) \cap D(i')| \leq 1$ for all different pairs $i, i' \in [n]$, $i \neq i'$.*

In Appendix C, we discuss how to construct an outer layer that has the sparse group caching structure, drawing on connections with coding theory and in particular cyclic and LDPC codes.

*Inner Layer:* In the inner layer, we consider each message group $g$ and the associated workers in $N(g)$ as a constrained pliable index coding instance, as shown in Fig. 3(b). Note that, given that we are interested in random shuffling, we regard the $c$-constraint to be a soft constraint and use a random coding scheme. We will discuss this at the end of Section IV-A. We then proceed as follows.

• Initialization: the cache of worker $i$ is filled with uniformly at random selected $m_1(1 - 1/r)$ messages from each group in $D(i)$, thus in total $s$ messages. This is done independently across workers.

• Iteration $t$: the master makes $m/m_1$ broadcast transmissions, one for every group. For each group $g$, the master selects uniformly at random $r$ messages in the group and transmits their linear combination, say $b_{j_1} + b_{j_2} + \ldots + b_{j_r}$. From the following Lemma 5, every worker in $N(g)$ can decode a new message with probability at least $1/e$. The workers who can decode a new message store it in their cache and discard an old message; they select the old message to discard uniformly at random from the messages in their cache that are also contained in the broadcast transmission, i.e., one from $\{b_{j_1}, b_{j_2}, \ldots, b_{j_r}\}$.

**Lemma 5.** *A worker with $m_1(1 - 1/r)$ cached messages from group $g$ that receives a linear combination $b_{j_1} + b_{j_2} + \ldots + b_{j_r}$ of $r$ messages uniformly at random selected from $g$, can decode a message it does not have with probability at least $1/e$.*

*Proof.* Without loss of generality, assume the worker has cached the messages $1, 2, \ldots, m_1(1 - 1/r)$ and requires a new message from the remaining $m_1/r$ messages. The probability that there is exactly one message in the $r$ data pieces $b_{j_1}, b_{j_2}, \ldots, b_{j_r}$ selected from the last $m_1/r$ data pieces is lower bounded by

$$
\begin{aligned}
p_1 &\triangleq \Pr\{\text{The worker can decode a new message}\} \\
&\geq \frac{\binom{m_1/r}{1}\binom{m_1(1-1/r)}{r-1}}{\binom{m_1}{r}} \\
&= \frac{(m_1-\frac{m_1}{r})(m_1-\frac{m_1}{r}-1)\ldots(m_1-\frac{m_1}{r}-r+2)}{(m_1-1)(m_1-2)\ldots(m_1-r+1)} \\
&\geq (\frac{m_1-\frac{m_1}{r}-r-2}{m_1-r+1})^{r-1} = (1-\frac{m_1-r}{r(m_1-r+1)})^{r-1} \\
&\geq (1-\frac{1}{r})^{r-1} \geq \frac{1}{e}
\end{aligned}
$$

$\square$

**Remark 1.** *Note that we can approximately analyze the inner layer performance using a random graph instance with probability $1/r$ that a message is in a client's request set.*

**Remark 2.** *We have two design parameters, $m_1$ and $r$. The first parameter $m_1$ is the size of each message group. Given a fixed $r$, we can see that when $m_1$ is smaller, the number of groups increases and the number of neighboring groups $|D(i)|$ also increases for a work node $i$, therefore, the number of broadcast transmissions increases and so does the Hamming distance between two states across iterations, namely $H(z_i^t, z_i^{t'})$. We will show this in Theorem 4. The second parameter $r$ represents the number of messages being encoded for making one transmissions, and $(1-1/r)$ represents the fraction of messages in a group being cached by an associated work node. If we increase $r$, we will expect an increase in communication efficiency, at the cost of a decrease in computational performance.*

**Remark 3.** *The proposed scheme satisfies a $c$-constraint in the following way. Assume that we require no more than $rc$ workers cache messages in group $g$, i.e., $|N(g)| \leq cr$ (or $\frac{ns}{m(r-1)} \leq c$). Hence, the constraint $c$ will be determined by the design parameter $r$. For example, for $c = 1$, then at most $r$ workers can be in $N(g)$, each one of them with $m_1(1-1/r)$ cached messages from this group. As at most $rc$ workers have cached messages from group $g$, from Lemma 5, we can see that on average at most $rcp_1$ (for some fixed $1 > p_1 \geq 1/e$) workers can update their cache with a new message during one transmission. Because we uniformly at random select which $r$ messages to encode, each message can be decoded by $cp_1$ workers on average. Hence, on average, the $c$-constraint is satisfied.*

## B. Algorithm Performance

We first formally define what we call a "random like" scheme, and then compare the communication cost with the index coding based method.

**Definition 4.** *Given a sparse group caching structure, we say that a data shuffling scheme is "random-like" if it satisfies the following two properties:*

• *Each worker $i$ uniformly at random caches $m_1(1 - 1/r)$ messages in any message group in $D(i)$.*

• *The workers' cache states are mutually independent at any iteration $t$; formally, for any $k \le n$ workers $i_1, i_2, \ldots, i_k$, the following is satisfied:*

$$\Pr\{z_{i_1}^t = v_1, z_{i_2}^t = v_2, \ldots, z_{i_k}^t = v_k\} = \Pr\{z_{i_1}^t = v_1\} \Pr\{z_{i_2}^t = v_2\} \ldots \Pr\{z_{i_k}^t = v_k\}, \quad (27)$$

The next theorem theoretically characterizes the performance of the proposed hierarchical data shuffling scheme.

**Theorem 4.** *The proposed hierarchical data shuffling scheme preserves the "random-like" property, and requires $G$ broadcast transmissions per iteration to achieve an average Hamming distance $H$ at least $\min\{\frac{2s}{em_1(1-1/r)}, 2(s - m_1 + m_1/r)\}$.*

*Proof.* From the proposed hierarchical data shuffling scheme, we can see that each message group makes one encoded broadcast transmission per iteration, resulting in $G$ broadcast transmissions per iteration.

Between the caches of any two workers the Hamming distance is at least $2(s - m_1 + m_1/r)$, since any two workers have common messages from at most one group.

Next, we evaluate the Hamming distance across iterations for the same worker. We denote by $z_i^t|_g$ the *truncated cache state* to indicate whether a message in group $g$ is stored by worker $i$ at iteration $t$, in other words, the subvector of the cache state $z_i^t$ only with elements corresponding to messages in group $g$. We first consider the average Hamming distance $H|_g$ only corresponding to the messages of a specific group $g$, i.e., the Hamming distance $H|_g \triangleq H(z_i^t|_g, z_i^{t'}|_g)$ of truncated cache states at two iterations $t$ and $t'$. The average Hamming distance across all iterations is at least the average Hamming Distance between two consecutive iterations (see Appendix D).

Hence, the average Hamming distance $H|_g$ can be lower bounded by:

$$H|_g \quad \geq 0 \cdot (1 - \tfrac{1}{e}) + 2 \cdot \tfrac{1}{e} = 2/e. \tag{28}$$

We then consider the average Hamming distance across all the groups in $D(i)$. Since $|D(i)| = \frac{s}{m_1(1-1/r)}$, this is at least $\frac{s}{m_1(1-1/r)} \frac{2}{e} = \frac{2s}{em_1(1-1/r)}$. Therefore, on average $H \geq \min\{\frac{2s}{em_1(1-1/r)}, 2(s - m_1 + m_1/r)\}$.

This scheme also allows us to maintain the randomness property for workers in $N(g)$ (see Appendix D), and therefore, the "random-like" property is preserved for all iterations. $\qquad \Box$

From Theorem 4, we can evaluate the benefits of our proposed scheme as compared to index coding based schemes. An index coding based scheme may require in the worst case $\Omega(n)$ broadcast transmissions and $\Theta(n/\log(n))$ for random graph instances to update one message in each cache, and thus $\Omega(ns/em_1(1-r))$ (in the worst case) and $\Theta(ns/em_1(1-1/r)\log(n))$ (for random graph instances) broadcast transmissions in each data shuffling iteration to guarantee a Hamming distance of $\frac{2s}{em_1(1-1/r)}$ across time. Using our proposed scheme, we need $m/m_1$ transmissions to achieve an average Hamming distance of $2s/em_1(1-1/r)$ across time. On average, each message is stored on $sn/m$ workers. The benefits of our proposed scheme over index coding (i.e., the ratio of the numbers of transmissions for index coding scheme and for our proposed scheme) is $O(sn/m)$ (in the worst case) and $O(\frac{sn}{m\log(n)})$ (for random graph instances). Additionally, finding the optimal index coding solution is NP-hard, while our scheme has linear complexity of encoding.

## V. Experimental Results

We conduct experiments on a distributed machine learning system for classification over a real dataset[4] [23]. The dataset contains data collected from a combined cycle power plant over 6 years (2006-2011), when the plant was set to work with full load. The four features represent temperature, ambient pressure, relative humidity, and exhaust vacuum. The goal is to train a classifier to categorize these data instances into 10 different classes distinguished by their net hourly electrical energy output. For a complete description of the dataset, please refer to [23].

We train the distributed classification model using a stochastic gradient descent method based on 500 training data instances (messages) and apply the model to the entire data set. We set the number of workers to $n = 20$ (with every two workers learning a class label) and the cache size to $s = 50$. We divide the messages into 50 groups, with 10 messages in each. We set the parameter $r = 2$, i.e., each worker has cached messages in 10 group. We carry out experiments 100 times by comparing our hierarchical pliable index coding based shuffling against: (i) no shuffling scheme without information exchange during learning performs local computations and combines the local learning models in the end; (ii) no shuffling scheme with information exchange during learning, i.e., carrying out step 2) and not 3) in the distributed computing process in Section II-A; and (iii) random shuffling with uniformly at random selected messages. For case (iii), once we randomly select what message to send to each worker, we use two approaches for broadcasting: uncoded broadcast transmissions, and index coding [2], [24]. We implemented index coding using the greedy graph coloring based heuristic approach in [24].

---

[4]https://archive.ics.uci.edu/ml/datasets/Combined+Cycle+Power+Plant

In Figs. 4 and 5 and Table. I, we compare the computational performance and communication cost of our pliable index coding based shuffling scheme with no shuffling and the random shuffling schemes. Note that the uncoded shuffling and index coding based shuffling use the same cached messages in each local computation, and only differ in the communication cost during data shuffling phase. We first observe that the no shuffling scheme without information exchange may suffer overfitting and the performance decreases in the learning process. We find that the no shuffling schemes achieve an error rate $9.2\%$ (without information exchange) and $5.1\%$ (with information exchange) higher than the random shuffling schemes on average (among the 100 experiments), and $16.5\%$ (without information exchange) and $10.8\%$ (with information exchange) in the worst case (among the 100 experiments). In contrast, our proposed pliable index coding based shuffling scheme achieves an error rate only $2.0\%$ (on average) and $5.2\%$ (in the worst case) higher than the random shuffling scheme. However, for the communication cost in terms of the number of broadcast transmissions, our proposed pliable index coding based semi-random shuffling scheme needs only $12\%$ (both on average and in the worst case) of broadcast transmissions required by the uncoded random shuffling scheme. In contrast, the index coding based random shuffling scheme requires at least $92\%$ of broadcast transmissions required by the uncoded random shuffling scheme. This indicates that, on average, our proposed pliable index coding based scheme saves $87\%$ of transmissions with a sacrifice of $2\%$ computational performance loss, compared with the index coding based random shuffling scheme.

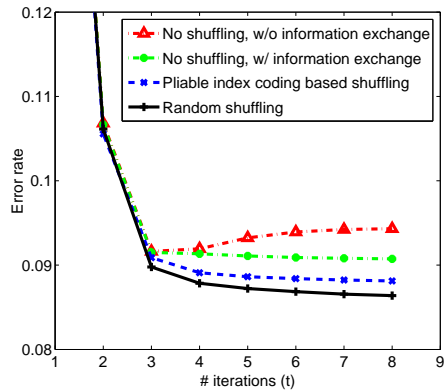

Fig. 4: Comparison of computational performance for data shuffling schemes. Experiments are carried out 100 times and the figure shows the average performance in terms of the error rate. The no shuffling schemes with or without information exchange do not make data shuffling during learning. The pliable index coding based scheme makes semi-random shuffling during iterations. The random shuffling schemes include both the uncoded transmission and the index coding based schemes, where the two schemes have cached the same data, but differ in communication cost.

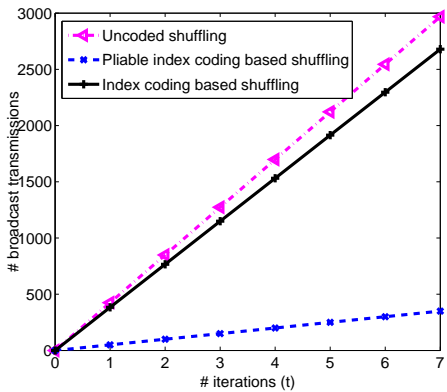

Fig. 5: Comparison of broadcast transmissions for data shuffling schemes. Experiments are carried out 100 times. The figure shows the average number of broadcast transmissions for the pliable index coding based shuffling scheme, the index coding based shuffling scheme, and the uncoded shuffling scheme.

## VI. CONCLUSION

In this paper, we presented a semi-random data shuffling scheme for distributed computing system, that balances the communication cost and computational performance. Our proposed scheme is based on pliable index coding framework and two modifications are made in order to achieve a good shuffling. One is to add data shuffling constraint that each message can go to at most a specific number of workers when the data are shuffled. The other is to reduce the

TABLE I: Performance comparison at the end of the 8-th iteration.

| | Average error rate over random shuffling | Worst case error rate over random shuffling | Average communication cost over uncoded random shuffling | Worst case communication cost over uncoded random shuffling |
|---|---|---|---|---|
| No shuffling, w/o information exchange | + 9.2% | +16.5% | - | - |
| No shuffling, w/ information exchange | + 5.1% | +10.8% | - | - |
| Pliable index coding based shuffling | +2.0% | +5.2% | -88.2% | -88.4% |
| Index coding based shuffling | - | - | -9.7% | -8.6% |

The average/worst case error rate over random shuffling scheme is calculated as $(r_A - r_{Rand})/r_{Rand}$, where $r_A$ is the average/worst case error rate for some scheme and $r_{Rand}$ is the average/worst case error rate for the random shuffling scheme. Similarly, the average/worst case communication cost over uncoded random shuffling is calculated as $(c_A - c_{Uncoded})/c_{Uncoded}$, where $c_A$ is the average/worst case number of broadcast transmissions for some scheme and $c_{Uncoded}$ is the average/worst case number of broadcast transmissions for the uncoded shuffling scheme.

correlation of cached content across iterations by a hierarchical data shuffling structure. Our results indicate potential benefits of our proposed scheme up to $O(ns/m)$ over index coding, where $ns/m$ is the average number of workers caching a message, and $m$, $n$, and $s$ are the numbers of messages, workers, and cache size, respectively. Experiments over real dataset show that our proposed pliable index coding based scheme saves $88\%$ of transmissions with a sacrifice of $2\%$ computational performance loss, compared with the index coding based random shuffling scheme.

# APPENDIX A

## PROOF OF THEOREM 1

In this appendix, we prove Theorem 1.

### A. Deciding if Optimal $L = 1$ is in P

We first show that deciding if the optimal code length equals $1$ is in P. To see this, we notice that if one transmission can make each client to receive a distinct message, then the server needs to encode exact $n$ messages for the transmission, one for each client. For a client $i$, if it can decode a message $b_j$, $j \in R_i$, then all other $n-1$ messages must be in its side information set following from the decoding criterion. Similarly, any one of the $n$ messages for encoding is in the side information set of $n-1$ clients and requested by the remaining one client. Hence, in the bipartite graph representation, if and only if we can find $n$ message vertices, such that each one has degree 1 and is connected to a different client vertex, then the optimal code length is 1. This can be tested by going over all message vertices, which runs in polynomial time.

## B. Deciding if Optimal $L = 2$ is NP-complete

We next show that deciding if optimal code length equals 2 is NP-complete. To prove this, we first introduce another NP-complete problem.

**Definition 5** (Distinct Labeling Problem). *We are given a universal set $U = \{1, 2, \ldots, u\}$ with $|U| = u$ elements, a fixed set of $\Pi$ labels $\{1, 2, \ldots, \Pi\}$, and a collection of size 3 subsets of $U$, i.e., $\mathcal{S} \subseteq 2^U$ and $|S| = 3$ for any $S \in \mathcal{S}$, where $2^U$ is the power set of $U$. The distinct labeling problem (DL) asks if we can label the elements using $\Pi$ labels such that every subset in $\mathcal{S}$ contains elements of 3 different labels. For short, we call it $\Pi$-DL problem for such a distinct labeling problem with $\Pi$ labels.*

**Lemma 6.** *$\Pi$-DL problem is NP-complete for $\Pi \geq 3$.*

*Proof.* It is easy to see that the $\Pi$-DL problem is in NP. We next show that we can use a polynomial time reduction from the graph coloring problem (a.k.a., chromatic number) to the $\Pi$-DL problem.

We reiterate the well-known decision version of graph coloring problem as follows [25]: it is NP-complete to decide whether the vertices of a given graph $G(V, E)$ can be colored using a fixed $\Pi \geq 3$ colors, such that no two neighboring vertices share the same color.

We perform the following mapping. We map each vertex in $V$ and each edge in $E$ as the universal set with $|U| = |V| + |E|$ elements. We map an edge $e \in E$ together with the two endpoints $x_1, x_2$ as a subset, where $e = \{x_1, x_2\}$. So there are in total $|\mathcal{S}| = |E|$ subsets.

We first show that if $G$ is $\Pi$-colorable, then we can find a solution for the $\Pi$-DL problem. We can assign a set $\{1, 2, \ldots, \Pi\}$ of colors to the vertices in $V$, such that no two neighboring vertices share a same color. When we map to the $\Pi$-DL problem, we notice that each edge appears in exact 1 subset, the one corresponding to this edge. Hence, we can use the following labeling scheme: label the elements corresponding to the vertices as the color used in $\{1, 2, \ldots, \Pi\}$; and label the edge using any label that is different from its two endpoints. This is a solution for the $\Pi$-DL problem.

On the other hand, if we have a solution for the $\Pi$-DL problem, we can find a solution for the graph coloring problem. We can label the elements corresponding to vertices using the $\Pi$ labels. Note that if two vertices $x_1$ and $x_2$ are adjacent to each other, i.e., $\{x_1, x_2\} \in E$, then according to the definition of $\Pi$-DL problem, these two elements $x_1$ and $x_2$ must have different labels. Hence, keep the labels of each vertex element, then we get a $\Pi$-coloring of the graph $G$.

□

We then prove that deciding if the optimal code length $L = 2$ is NP-complete for constrained pliable index coding problem over a finite filed $\mathbb{F}_q$.

First, we observe that we can decide if a given $2 \times m$ coding matrix $\boldsymbol{A}$ can satisfy a constrained pliable index coding instance from our decoding criterion. Indeed, given a coding matrix $\boldsymbol{A}$, one can list the messages a client can decode using the decoding criterion. Then we have a bipartite subgraph representation that has $n$ clients, some messages, and edges that connect each client with the message she can decode. We only need to check if the maximum matching in such a subgraph equals the number of clients $n$ using polynomial time. If and only if so, this coding matrix can satisfy the problem instance.

Next, we use a reduction from the $(q + 1)$-DL problem defined above to show that the constrained pliable index coding problem is NP-hard. We are given a $(q + 1)$-DL problem instance with the universal set $U = \{1, 2, \ldots, u\}$ and a collection of size 3 subsets $\mathcal{S} \subseteq 2^U$. We perform the following two mappings.

• For each subset, e.g., $S = \{x, y, z\} \in \mathcal{S}$ and $x, y, z \in U$, we map into a structure as show in Fig. 6. We map each element in the subset $S$ as a message vertex and add 3 client vertices $c_1, c_2, c_3$ in the constraint pliable index coding problem instance. We connect $c_1$ to $x$ and $y$, connect $c_2$ to $y$ and $z$, and connect $c_3$ to $z$ and $x$.

• For different subsets, if they contain the same element, we connect them using the following structure as shown in Fig. 7. For example, if the subsets $S_1 = \{x, y_1, z_1\}$, $S_2 = \{x, y_2, z_2\}$, and $S_3 = \{x, y_3, z_3\}$ all contain the element $x$, we connect a client vertex $c_x$ to all messages corresponding to $x$ and another additional message vertex $b_x$.

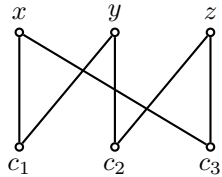

Fig. 6: Mapping a subset into a structure.

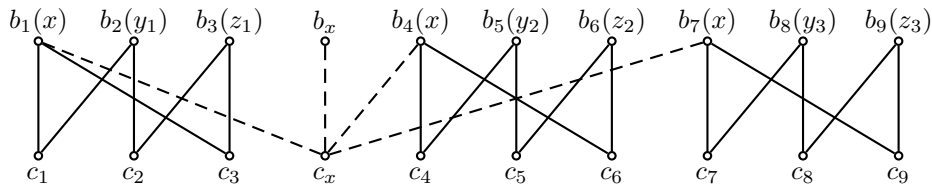

Fig. 7: Connecting the same elements in different subsets.

After this mapping, we can see that we construct a constrained pliable index coding instance with $n = 3|\mathcal{S}| + |U|$ clients and $m = n$ messages. We want to show that if and only if the $(q + 1)$-DL problem outputs a "Yes" answer, a code length 2 coding matrix can satisfy such a problem.

If for a "Yes" instance of $(q + 1)$-DL problem, we can find a labeling scheme using $q + 1$ labels to the elements. In finite field $\mathbb{F}_q$, we notice that the maximum number of vectors that are pair-wise independent is $q + 1$, e.g.,

$$\begin{bmatrix} 0 \\ 1 \end{bmatrix}, \begin{bmatrix} 1 \\ 0 \end{bmatrix}, \begin{bmatrix} 1 \\ 1 \end{bmatrix}, \begin{bmatrix} 1 \\ 2 \end{bmatrix}, \ldots, \begin{bmatrix} 1 \\ q - 1 \end{bmatrix}. \tag{29}$$

We consider each label as one of these $q+1$ vectors. Then for the coding matrix $\boldsymbol{A}$, we can assign the columns the same vector that correspond to the same element in subsets, e.g., $b_1$, $b_4$, and $b_7$ in Fig. 7. For the columns corresponding to an element not in subsets, e.g., $b_x$, we assign a different vector other than the one for the element in subsets. This is a valid coding matrix. Indeed, for 3 messages corresponding to a subset, e.g., $b_1$, $b_2$, and $b_3$, they are labeled using different labels from the $q + 1$-DL problem solution. Then, the 3 clients corresponding to this subset, i.e., $c_1$, $c_2$, and $c_3$, can decode $b_1$, $b_2$, and $b_3$, respectively, according to the decoding criterion. The client corresponding to an element not in subsets, e.g., $c_x$, can decode the corresponding message not in subsets, i.e., $b_x$, as coding vectors corresponding to messages $b_1$, $b_4$, and $b_7$ are the same and different from the coding vector corresponding to message $b_x$.

If for the constrained pliable index coding instance, a length 2 coding matrix $\boldsymbol{A}$ can make a successful shuffling. Then we notice that client $c_x$ should be satisfied by message $b_x$, since $b_x$ only connects to $c_x$ and $m = n$, which implies that each message needs to satisfy a client. In this case, the non-zero coding vector corresponding to $b_x$ is not in the space spanned by other coding vectors corresponding to $x$ in subsets, i.e., $b_1$, $b_4$, and $b_7$. As a result, the space spanned by coding vectors corresponding to the same element in subsets is a one dimensional space, e.g., the space spanned by coding vectors corresponding to $b_1$, $b_4$, and $b_7$. For clients and messages inside a subset, e.g., $b_1$, $b_2$, $b_3$, $c_1$, $c_2$, and $c_3$, there are two ways to satisfy these clients: one is $b_1$ to $c_1$, $b_2$ to $c_2$, $b_3$ to $c_3$; and the other one is $b_2$ to $c_1$, $b_3$ to $c_2$, $b_1$ to $c_3$. For both of these two ways, we notice that coding vectors corresponding to $b_1$, $b_2$, and $b_3$ should be pair-wise independent; otherwise, one of the clients cannot decode a new message, e.g., if coding vectors corresponding to $b_1$ and $b_2$ are dependent to each other, then $c_1$ cannot decode any new message. In addition, we observe that there are in total $q + 1$ 1-dimensional subspaces spanned by 2-dimensional vectors over finite filed $\mathbb{F}_q$, i.e., the spaces spanned by vectors in 29. Therefore, if we assign each space a label, then messages in the same subsets are using different labels and messages in different subsets corresponding to a same element are using the same label, resulting in a solution of the $q + 1$-DL problem.

## APPENDIX B

## PROOF OF LEMMA 4

In the following, we complete the proof of Lemma 4 for scenarios 2 and 3. The techniques used here are similar to that for scenario 1.

2) For scenario 2, we have

$$\sum_{B_2:B_2 \sim B_0} \Pr\{X_{B_2} = 1 | X_{B_0} = 1\} \leq \sum_{j=1}^{\mathbb{K}} \binom{\mathbb{K}}{j} \binom{m' - \mathbb{K}}{\mathbb{K} - j} \mathbb{K}! \frac{p^{\mathbb{K}}(1-p)^{\mathbb{K}(\mathbb{K}-1)}}{\bar{p}^{j\mathbb{K}}}. \tag{30}$$

Let us define the term inside the summation as $\Delta_j \triangleq \binom{\mathbb{K}}{j} \binom{m' - \mathbb{K}}{\mathbb{K} - j} \mathbb{K}! \frac{p^{\mathbb{K}}(1-p)^{\mathbb{K}(\mathbb{K}-1)}}{\bar{p}^{j\mathbb{K}}}$.

Then we can see that for $j = 1, 2, \ldots, \mathbb{K} - 1$, we have

$$
\begin{aligned}
\frac{\Delta_{j+1}}{\Delta_j} &= \frac{(\mathbb{K}-j)^2}{(j+1)(m'-2\mathbb{K}+j+1)} \bar{p}^{-\mathbb{K}} \\
&\leq \frac{\mathbb{K}^2}{m'} \bar{p}^{-\mathbb{K}} \\
&\leq \frac{\frac{1}{\log^2(1/\bar{p})} \log^2(m)}{m(1-o(1))} \frac{m \log^2(1/\bar{p})}{\log^3(m)} \\
&= o(1).
\end{aligned}
\tag{31}
$$

This implies that for all $j = 1, 2, \ldots, \mathbb{K}$, $\Delta_j \leq \Delta_1$.

For $\Delta_1$, we have the following

$$
\begin{aligned}
\frac{\Delta_1}{\mathbb{E}[Y_{\mathbb{K}}^{B'}]} &= \frac{\mathbb{K}\binom{m'-\mathbb{K}}{\mathbb{K}-1}\mathbb{K}! \frac{p^{\mathbb{K}}(1-p)^{\mathbb{K}(\mathbb{K}-1)}}{\bar{p}}}{\binom{m'}{\mathbb{K}}\binom{n'}{\mathbb{K}}\mathbb{K}! p^{\mathbb{K}}(1-p)^{\mathbb{K}(\mathbb{K}-1)}} \\
&= \frac{\mathbb{K}(m'-\mathbb{K})!(m'-\mathbb{K})!}{\bar{p}m'!(m'-2\mathbb{K}+1)!} \\
&\leq \frac{\mathbb{K}^2}{\bar{p}m(1-o(1))} \leq \frac{\log^2(m)(1+o(1))}{\bar{p}m\log^2(1/\bar{p})}.
\end{aligned}
\tag{32}
$$

Next, we have

$$
\mathbb{E}[|\mathcal{P}|] \leq \frac{1}{2}\mathbb{E}^2[Y_{\mathbb{K}}^{B'}]\mathbb{K}\frac{\log^2(m)(1+o(1))}{\bar{p}\log^2(1/\bar{p})m} \leq \mathbb{E}^2[Y_{\mathbb{K}}^{B'}]\frac{\log^3(m)}{2\bar{p}\log^3(1/\bar{p})m}
\tag{33}
$$

From Lemmas 2 and 3, we have $k_0^{B'} = \frac{1}{\log(1/(1-p))}[\log(\mathbb{K}) + \log(m - \mathbb{K}) - \log(\log(\mathbb{K}) + \log(m - \mathbb{K})) - \log\log(\frac{1}{1-p}) + \log(p)] + O(1) = \frac{1}{\log(1/(1-p))}[\log(\mathbb{K}) + \log(m - \mathbb{K}) - \log(\log(\mathbb{K}) + \log(m - \mathbb{K})) - \log\log(\frac{1}{1-p}) + \log(p)] + O(1) = \frac{1}{\log(1/(1-p))}[\log(m) - 2\log\log(\frac{1}{1-p}) + \log(p)] + O(1)$ and $\mathbb{E}[Y_{k_0^{B'}-3}^{B'}] \geq (\frac{m\mathbb{K}}{e})^{3(1+o(1))}$. Obviously, we have $\mathbb{K} < k_1^{B'} = k_0^{B'} - 3$ and $\mathbb{E}[Y_{\mathbb{K}}^{B'}] \geq (\frac{m\mathbb{K}}{e})^{3(1+o(1))}$. By setting the probability $p^{\dagger} = \frac{\bar{p}\log^3(\frac{1}{\bar{p}})m(1-o(1))}{\mathbb{E}[Y_{\mathbb{K}}^{B'}]\log^3(m)} < 1$, we can bound the average number of $X$, $\mathbb{E}[X]$, by

$$
\mathbb{E}[X] \geq \frac{\bar{p}\log^3(\frac{1}{\bar{p}})m(1-o(1))}{2\log^3(m)}.
\tag{34}
$$

We then can bound the following probability using Azuma's inequality

$$
\begin{aligned}
\Pr\{B' \text{ contains no } \mathbb{K}\text{-pattern}\} &\leq \exp(-\frac{\bar{p}^2\log^6(1/\bar{p})m^2(1-o(1))}{8\log^6(m)m'n'}) \\
&\leq \exp(-\frac{\bar{p}^2\log^7(1/\bar{p})m(1-o(1))}{8\log^7(m)}).
\end{aligned}
\tag{35}
$$

Therefore, we can bound the probability that any subgraph $B'$ induced by $m'$ messages and

$n'$ clients does not contain a $\mathbb{K}$-pattern:

$$
\begin{aligned}
\Pr\{\exists B' \in \mathcal{B}(B, m', n'), s.t., B' \text{ contains no } \mathbb{K}\text{-pattern}\} \ &\leq \binom{m}{m'}\binom{n}{n'}\exp(-\tfrac{\bar{p}^2 \log^7(1/\bar{p})m(1-o(1))}{8\log^7(m)}) \\
&\leq m^n 2^n \exp(-\tfrac{\bar{p}^2 \log^7(1/\bar{p})m(1-o(1))}{8\log^7(m)}) = o(1),
\end{aligned}
\tag{36}
$$

where the last equality follows from that $n \leq \log^{15}(m)$.

3) For scenario 3, we have

$$
\sum_{B_2:B_2\sim B_0}\Pr\{X_{B_2}=1|X_{B_0}=1\} \leq \sum_{j=1}^{\mathbb{K}}\sum_{i=1}^{\mathbb{K}c}\binom{\mathbb{K}}{j}\binom{m'-\mathbb{K}}{\mathbb{K}-j}\binom{\mathbb{K}c}{i}\binom{n'-\mathbb{K}c}{\mathbb{K}c-i}\binom{\mathbb{K}c}{c,c,\dots,c}\frac{p^{\mathbb{K}c(1-p)^{\mathbb{K}c(\mathbb{K}-1)}}}{\bar{p}^{ij}},
\tag{37}
$$

Let us define the term inside the summation as $\Delta_{i,j} \triangleq \binom{\mathbb{K}}{j}\binom{m'-\mathbb{K}}{\mathbb{K}-j}\binom{\mathbb{K}c}{i}\binom{n'-\mathbb{K}c}{\mathbb{K}c-i}\binom{\mathbb{K}c}{c,c,\dots,c}\frac{p^{\mathbb{K}c(1-p)^{\mathbb{K}c(\mathbb{K}-1)}}}{\bar{p}^{ij}}$.

Then we can see that for $j=1,2,\dots,\mathbb{K}$ and $i=1,2,\dots,\mathbb{K}c-1$, we have

$$
\begin{aligned}
\frac{\Delta_{i+1,j}}{\Delta_{i,j}} &= \frac{(\mathbb{K}c-i)^2}{(i+1)(n'-2\mathbb{K}c+i+1)}\bar{p}^{-j} \\
&\leq \frac{\mathbb{K}^2 c^2}{n'}\bar{p}^{-\mathbb{K}} \\
&\leq \frac{\frac{1}{\log^2(1/\bar{p})}c^3\log(n)}{n}\frac{n\log^2(1/\bar{p})}{c^3\log^3(n)} \\
&\leq 1.
\end{aligned}
\tag{38}
$$

This implies that for all $i=1,2,\dots,\mathbb{K}c$, $\Delta_{i,j} \leq \Delta_{1,j}$.

We also note that for $j=1,2,\dots,\mathbb{K}-1$, we have

$$
\begin{aligned}
\frac{\Delta_{1,j+1}}{\Delta_{1,j}} &= \frac{(\mathbb{K}-j)^2}{(j+1)(m'-2\mathbb{K}+j+1)}\bar{p}^{-1} \\
&\leq \frac{\mathbb{K}^2}{2\bar{p}(m'-2\mathbb{K}+2)} \\
&\leq \frac{\mathbb{K}^2}{\bar{p}m'} \\
&\leq \frac{\log^3(n)}{\bar{p}\log^2(1/\bar{p})m} \\
&= o(1).
\end{aligned}
\tag{39}
$$

This implies that for all $i=1,2,\dots,\mathbb{K}c$ and $j=1,2,\dots,\mathbb{K}$, $\Delta_{i,j} \leq \Delta_{1,j} \leq \Delta_{1,1}$.

For $\Delta_{1,1}$, we have the following

$$
\begin{aligned}
\frac{\Delta_{1,1}}{\mathbb{E}[Y_{\mathbb{K}}^{B'}]} &= \frac{\mathbb{K}\binom{m'-\mathbb{K}}{\mathbb{K}-1}\mathbb{K}c\binom{n'-\mathbb{K}c}{\mathbb{K}c-1}\binom{\mathbb{K}c}{c,c,\dots,c}\frac{p^{\mathbb{K}c(1-p)^{\mathbb{K}c(\mathbb{K}-1)}}}{\bar{p}}}{\binom{m'}{\mathbb{K}c}\binom{n'}{\mathbb{K}c}\binom{\mathbb{K}c}{c,c,\dots,c}p^{\mathbb{K}c}(1-p)^{\mathbb{K}c(\mathbb{K}-1)}} \\
&\leq \frac{\mathbb{K}^4 c^2}{\bar{p}m'n'} \\
&\leq \frac{c^3\log^6(n)}{\bar{p}mn\log^4(1/\bar{p})}.
\end{aligned}
\tag{40}
$$

Next, we have

$$\mathbb{E}[|\mathcal{P}|] \quad \leq \tfrac{1}{2}\mathbb{E}^2[Y_{\mathbb{K}}^{B'}]\mathbb{K}^2 c \frac{c^3 \log^6(n)}{\bar{p}mn\log^4(1/\bar{p})} \leq \mathbb{E}^2[Y_{\mathbb{K}}^{B'}]\frac{c^4 \log^8(n)}{2\bar{p}\log^6(1/\bar{p})mn} \tag{41}$$

From Lemmas 2 and 3, we have $k_0^{B'} = \frac{1}{\log(1/(1-p))}[\log(n) + \frac{\log(m)}{c} - 2\log(c) - (1+1/c)\log\log(n) - \frac{\log[\log(n)+\log(m)/c - 2\log(c)]}{c} - \frac{\log\log(1/(1-p))}{c} + \log(p)] + O(1)$ and $\mathbb{E}[Y_{k_0^{B'}-3}^{B'}] \geq \left(\frac{n}{ec^2\log(n)}\right)^{3c(1+o(1))}\left(\frac{m}{\log(n)}\right)^{3(1+o(1))}$. Obviously, we have $\mathbb{K} < k_1^{B'} = k_0^{B'} - 3$ and $\mathbb{E}[Y_{\mathbb{K}}^{B'}] \geq \left(\frac{n}{ec^2\log(n)}\right)^{3c(1+o(1))}\left(\frac{m}{\log(n)}\right)^{3(1+o(1))}$. By setting the probability $p^\dagger = \frac{\bar{p}\log^6(\frac{1}{\bar{p}})mn}{\mathbb{E}[Y_{\mathbb{K}}^{B'}]c^4\log^8(n)} < 1$, we can bound the average number of $X$, $\mathbb{E}[X]$, by

$$\mathbb{E}[X] \quad \geq \frac{\bar{p}\log^6(\frac{1}{\bar{p}})mn}{2c^4\log^8(n)}. \tag{42}$$

We then can bound the following probability using Azuma's inequality

$$\Pr\{B' \text{ contains no } \mathbb{K}\text{-pattern}\} \quad \leq \exp(-\frac{\bar{p}^2\log^{12}(1/\bar{p})m^2n^2}{8c^8\log^{16}(n)m'n'}) \\ \leq \exp(-\frac{\bar{p}^2\log^{12}(1/\bar{p})mn}{8c^7\log^{14}(n)}). \tag{43}$$

Therefore, we can bound the probability that any subgraph $B'$ induced by $m'$ messages and $n'$ clients does not contain a $\mathbb{K}$-pattern:

$$\Pr\{\exists B' \in \mathcal{B}(B, m', n'), s.t., B' \text{ contains no } \mathbb{K}\text{-pattern}\} \quad \leq \binom{m}{m'}\binom{n}{n'}\exp(-\frac{\bar{p}^2\log^{12}(1/\bar{p})mn}{8c^7\log^{14}(n)}) \\ \leq 2^{m+n}\exp(-\frac{\bar{p}^2\log^{12}(1/\bar{p})mn}{8c^7\log^{14}(n)}) = o(1), \tag{44}$$

where the last equality follows from that $c = o(\frac{n^{1/7}}{\log^2(n)})$.

## Appendix C

### Construction of Outer Layer Biregular Graph

In this appendix, we discuss in detail about the construction of outer layer biregular graph. Note that, this is equivalent to construct a $C_4$-free biregular graph (i.e., no 4-cycle as a subgraph). However, the fundamental understanding about this is still an open question in extreme graph theory. Here, we present three directions regarding constructing the outer layer biregular graph.

We use a biadjacency matrix[5] $B \in \{0, 1\}^{n \times m/m_1}$ to represent the bipartite graph, where the rows correspond to $n$ worker nodes and the columns correspond to $m/m_1$ message groups.

---

[5]For a bipartite graph $G(U \cup V, E)$, the biadjacency matrix is a $(0, 1)$ matrix of size $|U| \times |V|$, whose $(i, j)$ element equals 1 if and only if $i$ connects $j$.

According to our proposed outer layer architecture, $B$ is a matrix with equal number of $d_1$ 1s in each row and equal number of $d_2$ 1s in each column and no $2 \times 2$ submatrix of $B$ has all 1s.

• Construction through binary constant weight cyclic codes. Our first idea is to construct this matrix $B$ by stacking all codewords of a binary constant weight cyclic code [26], i.e., each codeword is a row of $B$, that has weight $d_1$, code length $m/m_1$, minimum Hamming distance $2(d_1 - 1)$ and number of codewords $n$.

Since the minimum Hamming distance is $2(d_1 - 1)$ for a constant weight $d_1$ code, any pair of the codewords has at most 1 position of overlapping of 1. Also for cyclic codes, the right or left shifting of each codeword is still a codeword. Hence, each column of $B$ goes through the same number of 1s and 0s with a different order. Therefore, the constructed matrix $B$ is a $C_4$-free biregular graph.

• A recursive construction algorithm, where we are inspired by ideas of how Gallager constructed the LDPC parity check matrix. Assume $m/m_1$ is divisible by $d_1$ and can be factorized as $m/m_1 = d_1 k_1 k_2 \ldots k_l$ for some primes $k_1, k_2, \ldots, k_l$. We then have the following recursive construction algorithm to construct a $n \times m/m_1$ $C_4$-free biregular matrix for $n = \frac{m}{m_1 d_1} i_1 i_2 \ldots i_l = k_1 i_1 k_2 i_2 \ldots k_l i_l$ with integers $i_1 \in [k_1], i_2 \in [k_2], \ldots, i_l \in [k_l]$. For convenience, we define $k_0 = i_0 = 1$.

Next, we show that we can recursively construct $C_4$ free bireguar matrices of size $k_1 i_1 k_2 i_2 \ldots k_{l'} i_{l'}$ by $d_1 k_1 k_2 \ldots k_{l'}$ for steps $l' = 0, 1, 2, \ldots, l$ that have weights $d_1$ for each row and $i_1 i_2 \ldots i_{l'}$ for each column.

− Initial step $l = 0$: we construct a matrix with $k_0 i_0 = 1$ row and $d_1 k_0 = d_1$ columns. We just simply construct a $1 \times d_1$ matrix of all 1s.

− Recursive step $l'$ to $l' + 1$: Assume that we have successfully construct a $C_4$ free bireguar matrix $B_1$ with $k_1 i_1 \ldots k_{l'} i_{l'}$ rows and $d_1 k_1 \ldots k_{l'}$ columns, with weights $d_1$ for each row. Now, we would like to show that we can construct another matrix $B_2$ with $k_1 i_1 \ldots k_{l'+1} i_{l'+1}$ rows and $d_1 k_1 \ldots k_{l'+1}$ columns that is also $C_4$ free bireguar. To realize this, we first construct a set of $i_{l'+1}$ matrices $B_{2,0}, B_{2,1}, \ldots, B_{2,i_{l'+1}-1}$, each with size $k_1 i_1 \ldots k_{l'} i_{l'} k_{l'+1}$ by $d_1 k_1 \ldots k_{l'+1}$. Then we stack these matrices together to get $B_2$.

To get these matrices $B_{2,i}$ ($i = 0, 1, \ldots, i_{l'+1} - 1$) from $B_1$, we replace the 1s in $B_1$ with a $k_{l'+1} \times k_{l'+1}$ identity matrix or its circularly shifted version and the 0s with a $k_{l'+1} \times k_{l'+1}$ all 0 matrix.

In particular, let us denote by $P$ the $k_{l'+1} \times k_{l'+1}$ cyclic permutation matrix, i.e., 1-left circularly shifted version of the identity matrix, shown as follows:

$$
\begin{bmatrix}
0 & 0 & \ldots & 0 & 1 \\
1 & 0 & \ldots & 0 & 0 \\
0 & \ddots & \ddots & \vdots & \vdots \\
\vdots & \ddots & \ddots & 0 & 0 \\
0 & \ldots & 0 & 1 & 0
\end{bmatrix}.
\tag{45}
$$

Using this matrix $P$, we can use a simple multiplication $P^i$ to represent the $i$-left circularly shifted version of the identity matrix.

We then have the following $k_{l'+1}$ ways to form matrices $B_{2,i}$:

(1) Replace $d_1$ 1s in each row of $B_1$ by $I, I, I, \ldots, I$. Let us denote this matrix by $B_{2,0}$.

(2) Replace $d_1$ 1s in each row of $B_1$ by $I, P, P^2, \ldots, P^{k_{l'+1}-1}$. Let us denote this matrix by $B_{2,1}$.

(3) Replace $d_1$ 1s in each row of $B_1$ by $I, P^2, P^4, \ldots, P^{2(k_{l'+1}-1)}$. Let us denote this matrix by $B_{2,2}$.

(4) Replace $d_1$ 1s in each row of $B_1$ by $I, P^i, P^{2i}, \ldots, P^{i(k_{l'+1}-1)}$. Let us denote these matrices by $B_{2,i}$ for $i = 3, 4, \ldots, k_{l'+1} - 1$.

The matrix $B_2$ is formed by arbitrarily stacking $i_{l'+1}$ of the above $k_{l'+1}$ matrices together (for simplicity, we select the first $i_{l'+1}$ of them). For the construction, it is not hard to see that the constructed matrix $B_2$ is $C_4$-free biregular.

Indeed, we can use an induction on $l'$ to show that the constructed matrix is $C_4$-free biregular. The initial condition $l' = 0$ holds according to our algorithm. Assume the matrix $B_1$ holds for $l'$, then we consider the matrix $B_2$ for $l' + 1$. Since the matrix $B_1$ is biregular, then each basic matrix $B_{2,i}$ is biregular and the stacked matrix $B_2$ is biregular. We focus on proving the $C_4$-free. First note that the matrix $B_{2,i}$ is $C_4$-free according to the construction and the assumption that $B_1$ is $C_4$-free. Then if we can find a $2 \times 2$ submatrix in $B_{2,i}$ and $B2, i'$ for some $i \neq i'$ and $j \neq j'$:

$$
\begin{bmatrix}
P^{ij} & P^{ij'} \\
P^{i'j} & P^{i'j'},
\end{bmatrix}
\tag{46}
$$

which requires that $i'j - ij = i'j' - ij'$ or $(i' - i)(j' - j) = 0$, resulting in a contradiction. Therefore, $B_2$ is also $C_4$-free.

• A third idea is to relax the constraint that $B$ is a $C_4$-free biregular graph. As we may see that even if two worker nodes have cached data pieces in slightly more than one group, it is still acceptable. Also it is not crucial if the degrees of each group vertex changes slightly. Therefore, a simple randomized algorithm would work by randomly selecting $n$ subsets among the $\binom{m/m_1}{d_1}$ subsets of size $d_1$ groups.

## Appendix D

### Properties of Pliable Index Coding Based Shuffling

#### A. Hamming Distance Analysis

We analyze the Hamming distance of our pliable index coding based shuffling. We first note that across different worker nodes, the Hamming distance is at least $2(s - m_1 + m_1/r)$, as in the outer layer of the transmission structure, two different worker nodes have common messages in no more than one group.

Next, we evaluate the Hamming distance across iterations for the same worker $i$. Let us define a *truncated cache state* on group $g$ for worker $i$ at iteration $t$, $z_i^t|_g \in \{0, 1\}^{m_1}$, as a $m_1$-tuple that consists of coordinates of $z_i^t$ corresponding to messages in group $g$. We first consider the Hamming distance $H|_g$ between truncated cache state on a specific group $g$ for worker $i$ across iterations. We claim that the average Hamming distance $H|_g$ across all iterations is at least the average Hamming Distance between two consecutive iterations, i.e., for two given iterations $t_1 < t_2$, $\Pr\{z_i^{t_1}|_g = z_i^{t_2}|_g\} \leq \Pr\{z_i^{t_1}|_g = z_i^{t_1+1}|_g\}$.

To prove this, we use a random walk model on a graph $G(V, E)$ that is constructed as follows. Each vertex $v \in V$ corresponds to one of $\binom{m_1}{m_1(1-1/r)}$ possible *truncated cache states* $z_i^t|_g$, or state for short, i.e., all binary vectors of length $m_1$ and weight $m_1(1 - 1/r)$. There is an edge between two states $v_1$ and $v_2$ if and only if their Hamming distance is no more than 2, i.e., each vertex $v$ has a self-loop and there is an edge connecting two vertices of Hamming distance 2. Thus, a vertex $v$ has $m_1^2(1/r - 1/r^2)$ connections with other vertices. Originally, worker $i$ is in any of the $\binom{m_1}{m_1(1-1/r)}$ possible states with equal probability. Using our proposed shuffling scheme, after each iteration, worker $i$ remains in the same state with probability $1 - p_1 \leq 1 - 1/e$ ($p_1$ is defined as the probability that a worker can decode a new message during each transmission) and changes to a neighboring state with probability $\frac{p_1}{em_1^2(1/r - 1/r^2)}$ according to Lemma 5. Assume at iterations $t_1$, worker $i$ is in some state $v_1 \in V$. At iteration $t_2$, worker $i$'s state is a random variable with some distribution. Let us denote by $p_v^t$ the probability that worker $i$ is in state $v$ at iteration $t$. Then we have the flow conservation equation:

$$
\begin{aligned}
p_{v_1}^{t_2} &= p_{v_1}^{t_2-1}(1-p_1) + \sum_{v \neq v_1 : \{v,v_1\} \in E} p_v^{t_2-1} \frac{p_1}{m_1^2(1/r-1/r^2)} \\
&= p_{v_1}^{t_2-1}(1-p_1) + p_{v_0}^{t_2-1} \frac{p_1}{m_1^2(1/r-1/r^2)} m_1^2(1/r-1/r^2) \\
&\leq p_{v_1}^{t_2-1}(1-p_1) + \frac{1-p_{v_1}^{t_2-1}}{m_1^2(1/r-1/r^2)} p_1 \\
&\leq 1-p_1 = p_{v_1}^{t_1+1},
\end{aligned}
\tag{47}
$$

where the second equality holds because the probabilities for worker $i$ in $v_1$'s neighbors, $p_v^{t_2-1}$ for $v \neq v_1 : \{v,v_1\} \in E$, are all equal by symmetry, and thus we can pick a fixed neighbor $v_0$ of $v_1$; the first inequality holds because $p_{v_0}^{t_2-1}$ is at most $\frac{1-p_{v_1}^{t_2-1}}{m_1^2(1/r-1/r^2)}$, i.e., worker $i$ has equal probability in any of $v_1$'s neighbors by symmetry and the probability that worker $i$ is in one of $v_1$'s neighbors is at most $1 - p_{v_1}^{t_2-1}$; and the second inequality holds because the function $g(p_{v_1}^{t_2-1}) = p_{v_1}^{t_2-1}(1-p_1) + \frac{1-p_{v_1}^{t_2-1}}{m_1^2(1/r-1/r^2)} p_1$ is an increasing function and achieves the maximum for $p_{v_1}^{t_2-1} = 1$. Our claim is proved.

Hence, the average Hamming distance $H|_g$ can be lower bounded by:

$$
H|_g \geq 0 \cdot (1-\tfrac{1}{e}) + 2 \cdot \tfrac{1}{e} = 2/e.
\tag{48}
$$

We then consider the average Hamming distance across all the groups in $D(i)$. Since $|D(i)| = \frac{s}{m_1(1-1/r)}$, this is at least $\frac{s}{m_1(1-1/r)} 2/e = \frac{2s}{em_1(1-1/r)}$. Therefore, on average $H \geq \min\{\frac{2s}{em_1(1-1/r)}, 2(s-m_1+m_1/r)\}$.

## B. Independence and Randomness Preserving Property

Originally, if the worker nodes in $N(g)$ have independently and uniformly at random cached $m_1(1-1/r)$ messages in group $g$, then we observe that the pliable index coding based shuffling scheme maintains this "independence and randomness" property. Without loss of generality, assume the worker nodes in $N(g)$ are $1, 2, \ldots, n_1$, where $n_1 = |N(g)|$. Again, we use the graph constructed above.

**Corollary 1.** *The pliable index coding based shuffling scheme maintains the "independence and*

*randomness" property. Formally, if the following two properties hold for iteration $t$:*

$$\Pr\{z_1^t|_g = v_1, z_2^t|_g = v_2, \ldots, z_{n_1}^t|_g = v_{n_1}\} = \Pr\{z_1^t|_g = v_1\}\Pr\{z_2^t|_g = v_2\}\ldots\Pr\{z_{n_1}^t|_g = v_{n_1}\},$$
(49)

*for any state tuple $(v_1, v_2, \ldots, v_{n_1}) \in V^{n_1}$, and*

$$\Pr\{z_i^t|_g = v_i\} = \frac{1}{|V|},$$
(50)

*for any worker $i \in [n_1]$ and state $v_i \in V$; then these two properties also hold for iteration $t+1$:*

$$\Pr\{z_1^{t+1}|_g = v_1', z_2^{t+1}|_g = v_2', \ldots, z_{n_1}^{t+1}|_g = v_{n_1}'\}$$
$$= \Pr\{z_1^{t+1}|_g = v_1'\}\Pr\{z_2^{t+1}|_g = v_2'\}\ldots\Pr\{z_{n_1}^{t+1}|_g = v_{n_1}'\},$$
(51)

*for any state tuple $(v_1', v_2', \ldots, v_{n_1}') \in V^{n_1}$, and*

$$\Pr\{z_i^{t+1}|_g = v_i'\} = \frac{1}{|V|},$$
(52)

*for any worker $i \in [n_1]$ and state $v_i \in V$.*

*Proof.* The second property is obvious. Indeed, by symmetry of the constructed graph, if worker $i$ is in every state with equal probability, then after one iteration (one random walk), worker $i$ remains in every state with equal probability.

We then show the first property. We have the following

$$\Pr\{z_1^{t+1}|_g = v_1', \ldots, z_{n_1}^{t+1}|_g = v_{n_1}'\}$$
$$= \sum_{(v_1, \ldots, v_{n_1})} \Pr\{z_1^t|_g = v_1, \ldots, z_{n_1}^t|_g = v_{n_1}\} \cdot$$
$$\qquad\qquad \Pr\{z_1^{t+1}|_g = v_1', \ldots, z_{n_1}^{t+1}|_g = v_{n_1}' \Big| z_1^t|_g = v_1, \ldots, z_{n_1}^t|_g = v_{n_1}\}$$
$$= \frac{n_1}{|V|} \sum_{(v_1, \ldots, v_{n_1})} \Pr\{z_1^{t+1}|_g = v_1, \ldots, z_{n_1}^{t+1}|_g = v_{n_1} \Big| z_1^t|_g = v_1', \ldots, z_{n_1}^t|_g = v_{n_1}'\}$$
$$= \frac{n_1}{|V|},$$
(53)

where the first equality holds due to the total probability theorem; the second equality holds because of the initial two properties for iteration $t$, i.e., e.q. (49) and (50), and the "reversibility property" of the random walk, i.e.,

$$\Pr\{z_1^{t+1}|_g = v_1', \ldots, z_{n_1}^{t+1}|_g = v_{n_1}' \Big| z_1^t|_g = v_1, \ldots, z_{n_1}^t|_g = v_{n_1}\}$$
$$= \Pr\{z_1^{t+1}|_g = v_1, \ldots, z_{n_1}^{t+1}|_g = v_{n_1} \Big| z_1^t|_g = v_1', \ldots, z_{n_1}^t|_g = v_{n_1}'\}.$$
(54)

The "reversibility property" describes that the probability walking from $(v_1, \ldots, v_{n_1})$ to $(v'_1, \ldots, v'_{n_1})$ is equal to that of walking from $(v'_1, \ldots, v'_{n_1})$ to $(v_1, \ldots, v_{n_1})$. Indeed, if we use the same coded transmission and a reverse discarding process, then we achieve the goal. For example, if a worker has messages $\{1, 2, 3\}$ in its cache, and the transmission is $b_1 + b_2 + b_3 + b_4$; then the worker decodes message $4$ and replaces message $1$ and at last has cached messages $\{2, 3, 4\}$. If we reverse the process, we start from messages $\{2, 3, 4\}$ in cache; using the same transmission $b_1 + b_2 + b_3 + b_4$, the worker decodes $b_1$ and replaces message $4$, resulting in cached messages $\{1, 2, 3\}$. This can be done with equal probability across all workers. The corollary is proved. $\quad\square$